\documentclass[journal=jpclcd,manuscript=letter]{achemso}

\usepackage[version=3]{mhchem} % Formula subscripts using \ce{}
\newcommand{\beginsupplement}{%
        \setcounter{table}{0}
        \renewcommand{\thetable}{S\arabic{table}}%
        \setcounter{figure}{0}
        \renewcommand{\thefigure}{S\arabic{figure}}%
     }

\author{Alexander Mikhaylov}
\affiliation[JILA]
{JILA, 440 UCB, University of Colorado, Boulder, CO 80305, USA}
\altaffiliation{A.M., R.N.W. and K.M.P. contributed equally to this paper preparation}

\author{Ryan N. Wilson}
\affiliation[JILA]
{JILA, 440 UCB, University of Colorado, Boulder, CO 80305, USA}
\alsoaffiliation[PhysDept]
{Department of Physics, 390 UCB, University of Colorado, Boulder, CO 80305, USA}
\altaffiliation{A.M., R.N.W. and K.M.P. contributed equally to this paper preparation}

\author{Kristen M. Parzuchowski}
\affiliation[JILA]
{JILA, 440 UCB, University of Colorado, Boulder, CO 80305, USA}
\alsoaffiliation[PhysDept]
{Department of Physics, 390 UCB, University of Colorado, Boulder, CO 80305, USA}
\altaffiliation{A.M., R.N.W. and K.M.P. contributed equally to this paper preparation}

\author{Michael D. Mazurek}
\affiliation[NISTB]
{National Institute of Standards and Technology,  325 Broadway, Boulder, CO 80305, USA}

\author{Charles H. Camp Jr}
\affiliation[NIST1]
{National Institute of Standards and Technology, 100 Bureau Dr, Gaithersburg, MD 20899, USA}

\author{Martin J. Stevens}
\affiliation[NISTB]
{National Institute of Standards and Technology,  325 Broadway, Boulder, CO 80305, USA}

\author{Ralph Jimenez}
\affiliation[JILA]
{JILA, 440 UCB, University of Colorado, Boulder, CO 80305, USA}

\alsoaffiliation[DeptChem]
{Department of Chemistry,  215 UCB, University of Colorado, Boulder, CO 80305, USA}

\email{rjimenez@jila.colorado.edu}

\title[An \textsf{achemso} demo]
  {Hot-Band Absorption Can Mimic Entangled Two-Photon Absorption}

\abbreviations{C2PA, E2PA, E2PEF, C2PEF, HBA, SPDC}

\begin{document}

\setlength{\fboxrule}{0 pt}

\begin{tocentry}

%Some journals require a graphical entry for the Table of Contents.This should be laid out ``print ready'' so that the sizing of the text is correct. Inside the \texttt{tocentry} environment, the font used is Helvetica 8\,pt, as required by \emph{Journal of the American ChemicalSociety}. The surrounding frame is 9\,cm by 3.5\,cm, which is the maximum permitted for  \emph{Journal of the American Chemical Society} graphical table of content entries. The box will not resize if the content is too big: instead it will overflow the edge of the box. This box and the associated title will always be printed on a separate page at the end of the document.

\includegraphics[width=5cm,height=5cm]{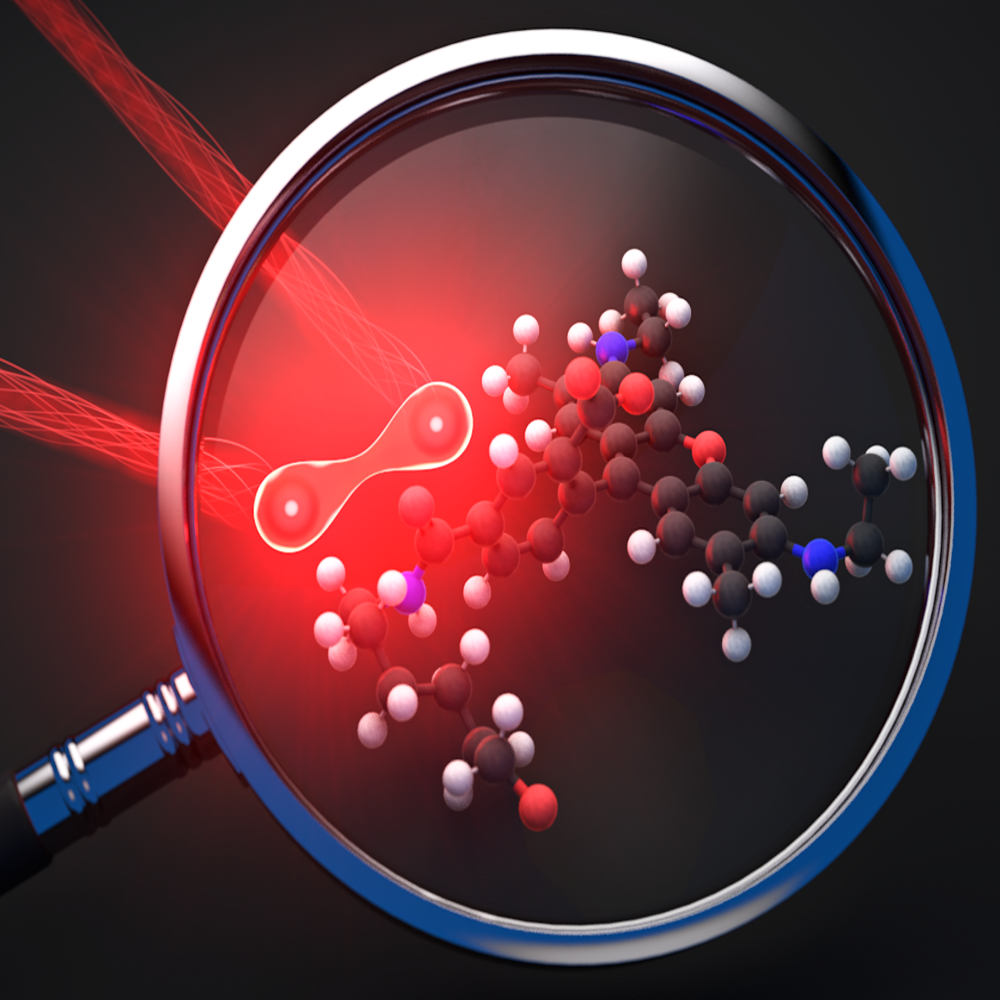}

\end{tocentry}

\begin{abstract}
 It has been proposed that entangled two-photon absorption (E2PA) can be observed with up to $10^{10}$ lower photon flux than its classical counterpart, therefore enabling ultra-low-power two-photon fluorescence microscopy. However, there is a significant controversy regarding the magnitude of this quantum enhancement in excitation efficiency. We investigated the fluorescence signals from Rhodamine 6G and LDS798 excited with a CW laser or an entangled photon pair source at $\approx$1060~nm. We observed a signal that originates from hot-band absorption (HBA), which is one-photon absorption from thermally-populated vibrational levels of the ground electronic state. This mechanism, which has not been previously discussed in the context of E2PA, produces a signal with a linear power dependence, as would be expected for E2PA. %However, unlike E2PEF, the signal decreases linearly rather than quadratically with attenuation of the entangled photons by linear losses.% 
 For the typical conditions under which E2PA measurements are performed, contributions from the HBA process could lead to a several orders-of-magnitude overestimate of the quantum advantage.
  %Aiming to investigate the enhancement of entangled two-photon absorption (E2PA) relative to coherent (classical) two-photon absorption (C2PA) in two fluorophores Rhodamine 6G and LDS798, we observe a one-photon fluorescence signal which seems to originate from absorption associated with the low energy vibronic states and is not related to E2PA. This fluorescence signal can be observed with both photon pairs and coherent excitation sources. This unexpected signal could lead to misleading conclusions regarding the magnitude of E2PA and thus to a many orders of magnitude overestimate of the advantage for E2PA over C2PA.
\end{abstract}


%\section{Introduction}

The implementation of non-classical light sources in spectroscopic and sensing methods has been a long-standing goal for advancing many practical applications of quantum science. One particularly intriguing possibility is to use a time-energy entangled photon pair source for two-photon absorption (2PA) excitation instead of a coherent, laser-based (classical) source. Here we refer to the latter regime as classical two-photon absorption (C2PA). It has been predicted that if entangled photons generated via spontaneous parametric down conversion (SPDC) are used for excitation then the resulting entangled two-photon absorption (E2PA) rate should scale linearly with the excitation flux and the process efficiency can be boosted relative to C2PA at low photon flux\cite{1989Banacloche,1990Javanainen}. Entangled photon excitation might therefore enable ultra-low-power two-photon excited fluorescence imaging, which would be particularly advantageous for limiting perturbation and damage of fragile biological samples.   
The favorable scaling behavior stems from the linear dependence of the 2PA rate on the second order correlation function, $g^{(2)}$ \cite{1968Mollow,2021Parzuchowski}. In addition to this absorption efficiency enhancement from the photon statistics, further enhancement is possible from the spectral shape and bandwidth of the frequency anticorrelated photon pairs \cite{2020Raymer,2021Carnio}. Whether these mechanisms can provide a practical advantage for E2PA in molecules is still unclear. 

Since 2004, numerous publications have reported E2PA and entangled two-photon excited fluorescence (E2PEF) for many different chromophores, reporting large excitation efficiencies \cite{2004French,2006Lee,2009Harpham,2010Guzman,2013Upton,2017Varnavski,2018Monsalve,2020Schatz}. The resulting E2PA cross sections, $\sigma_{\text{E2PA}}$, may be as large as $10^{-17}$~$\text{cm}^2$, which is on the same order of magnitude as a moderately-strong one-photon absorption (1PA) transition. More recently, however, a number of studies have reported conflicting results which cast doubt on the large enhancements claimed in those reports \cite{2019Ashkenazy,2020Mikhaylov,2021Corona,2021Raymer,2021Landes}. For example, three different groups employed E2PEF measurements to determine the $\sigma_{\text{E2PA}}$ of Rhodamine 6G (Rh6G) \cite{2021Tabakaev, 2021Parzuchowski, 2020landes}. Tabakaev \textit{et al.}~\cite{2021Tabakaev} measured E2PEF using CW SPDC excitation at 1064~nm with up to $5\times10^{8}$~pairs/sec. Although several tests were performed to rule out one-photon mechanisms as the origin of the measured signal, the observed dependence on time delay was inconsistent with expectations \cite{2020Lavoie}. The authors concluded that $\sigma_{\text{E2PA}}$ was $(0.99-1.9)\times 10^{-21}$~$\text{cm}^2$ for a range of fluorophore concentrations. In a separate study, Parzuchowski \textit{et al.}~\cite{2021Parzuchowski} observed no E2PEF using a pulsed SPDC excitation source at 810~nm with approximately $9\times10^{9}$~photons/sec. This result was used to determine an upper bound on the cross section for Rh6G of $\sigma_{\text{E2PA}} = 1.2\times 10^{-25}$~$\text{cm}^2$, which is nearly four orders of magnitude smaller than the value reported by Tabakaev \textit{et al.}~\cite{2021Tabakaev}. The null result of Parzuchowski \textit{et al.} was supported by results from a study by Landes \textit{et al.}~\cite{2020landes}. In this case, a CW SPDC excitation source at 1064~nm with $2\times10^{9}$~pairs/s was used, along with dispersion control and sum frequency generation measurements to optimize the excitation radiation parameters and the signal collection efficiency. However no measurable E2PEF was observed~\cite{2020landes}. 
The origin of the $\approx$10,000-fold variation in reported $\sigma_{E2PA}$ values is unclear. 

\begin{figure}[h!]
    \centering
    \includegraphics[width=1\textwidth]{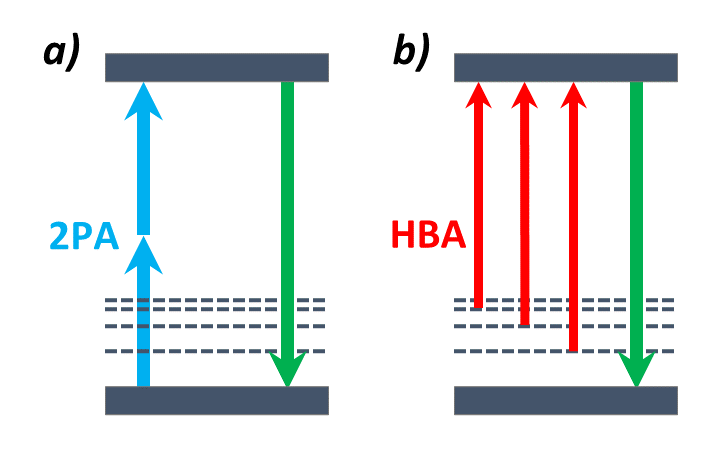}
    \caption{
    A schematic of 2PA (a) and HBA (b) with electronic and vibronic levels indicated by solid and dashed lines respectively. The 2PA excitation source (blue) may have high energy components (red) that resonate with 1PA (hot-band) transitions.
    Note that the fluorescence emitted by the two mechanisms (green arrows) is indistinguishable.}  
    \label{fig:HBA}
\end{figure}

Here we focus on hot-band absorption (HBA) which can contribute to signals measured with SPDC and mimic certain characteristics of E2PA. HBA is a classical 1PA process from the thermally-populated vibronic levels of the ground electronic state. 
Figure~\ref{fig:HBA} shows a schematic of a two electronic level system (solid grey lines) with vibronic levels of the ground state indicated by dashed grey lines. A 2PA transition is allowed between these two electronic states (blue vertical arrows, Fig.~\ref{fig:HBA}~\textit{a}). If the excitation source has a broad spectrum or is tuned far away from the peak of the ``0-0" transition, its radiation can stimulate transitions involving the vibronic manifold of the ground electronic state. If 1PA transitions between these levels and the upper electronic state are allowed, HBA may take place (red vertical arrows, Fig.~\ref{fig:HBA}~\textit{b}). Although the probability of HBA transitions is very low, C2PA is also inefficient, thus the magnitude of the signals from the two processes can be comparable under certain conditions. The system relaxes back to the ground electronic state emitting fluorescence photons (``anti-Stokes" emission; green vertical arrows), which are indistinguishable for the two mechanisms.  

HBA has been shown to play a crucial role in C2PA measurements~\cite{1983Apatin, 1995Okamura,2003Drobizhev,2007Makarov,2007Rebane,2008Starkey}, but has not been discussed in the E2PA literature. The importance of including HBA in the analysis of C2PA data has been detailed in a study by Drobizhev $\latin{et~  al.}$~\cite{2003Drobizhev}, where C2PA and HBA were simultaneously observed in a series of meso-tetra-alkynyl-porphyrins. In this study, the excitation frequency $\nu$ was detuned far to the red of the chromophore's ``0-0" transition frequency ($\nu_{\text{max}}$) to avoid direct one-photon excitation of the lowest energy transition. However, the detuning ($\nu_{\text{max}}-\nu$) was insufficient to avoid excitation from the vibronic manifold of the ground electronic state. Temperature-dependent measurements were conducted to decouple the roles of the two excitation pathways.

An additional complication arises in distinguishing HBA from E2PA using power dependence. Since HBA is a 1PA process, it scales linearly with excitation power. 
When the SPDC photon flux is sufficiently low that pairs are separated in time, the E2PA rate is also predicted to scale linearly with excitation power. However, when linear losses act on the produced pairs, E2PA exhibits the unique signature of scaling quadratically with attenuation of the SPDC beam. 
This behavior is also expected for other two-photon processes, as clearly demonstrated for sum frequency generation~\cite{2005Dayan}. Thus, to confirm the origin of a potential E2PA signal, both power dependencies should be measured. In earlier reports the linear dependence on the pump alone was taken as proof that the signals originated from E2PA. However this signature is consistent with many one-photon mechanisms~\cite{2021Parzuchowski}, including HBA. This amalgamation of signals, corrupting the purely quantum-enhanced 2PA signal, would lead to misleading conclusions regarding the efficiency of E2PA and its dependence on molecular properties and on the quantum state of the light.

Here we report 2PA measurements on Rh6G and LDS798 (CAS No 92479-59-9) dissolved in methanol and deuterated chloroform ($\text{CDCl}_{3}$), respectively. Rh6G is particularly interesting because it was studied in the prior E2PA reports mentioned above and has well known C2PA properties~\cite{2016deReguardatti}. LDS798 is another commercially-available fluorophore with a large C2PA cross section at 1064~nm~\cite{2011Makarov,2020Drobizhev}. According to a simple probabilistic model of the E2PA process proposed in Fei \textit{et al.}~\cite{1997Fei}, a large C2PA cross section implies a large E2PA cross section as well.
We use two CW sources operated near 1060~nm ---a laser and time-energy entangled photon pairs generated via SPDC---to independently excite the samples under identical conditions. We observe no measurable E2PEF signal from Rh6G with the maximum available SPDC power. In contrast, a signal is observed from the LDS798 sample. Upon further investigation, we find this signal does not show the excitation power scaling characteristics of E2PA. We attribute this fluorescence signal to HBA. Temperature-dependent measurements, excitation wavelength-dependent measurements, and modelling of the signals support our conclusions. We propose that HBA may be responsible for absorption signals observed with SPDC excitation. We emphasize the importance of including additional verification tests to elucidate the origin of signals measured with SPDC excitation. 

%%%%%%%%%%%%%%%%%%%%%%%%%%%%%%%%%%%%%%%%%%%%%%%%%%%%%%%%%%%%

%\section{Results and discussion}

\begin{figure}[h!]
    \centering
    \includegraphics[width=1\textwidth]{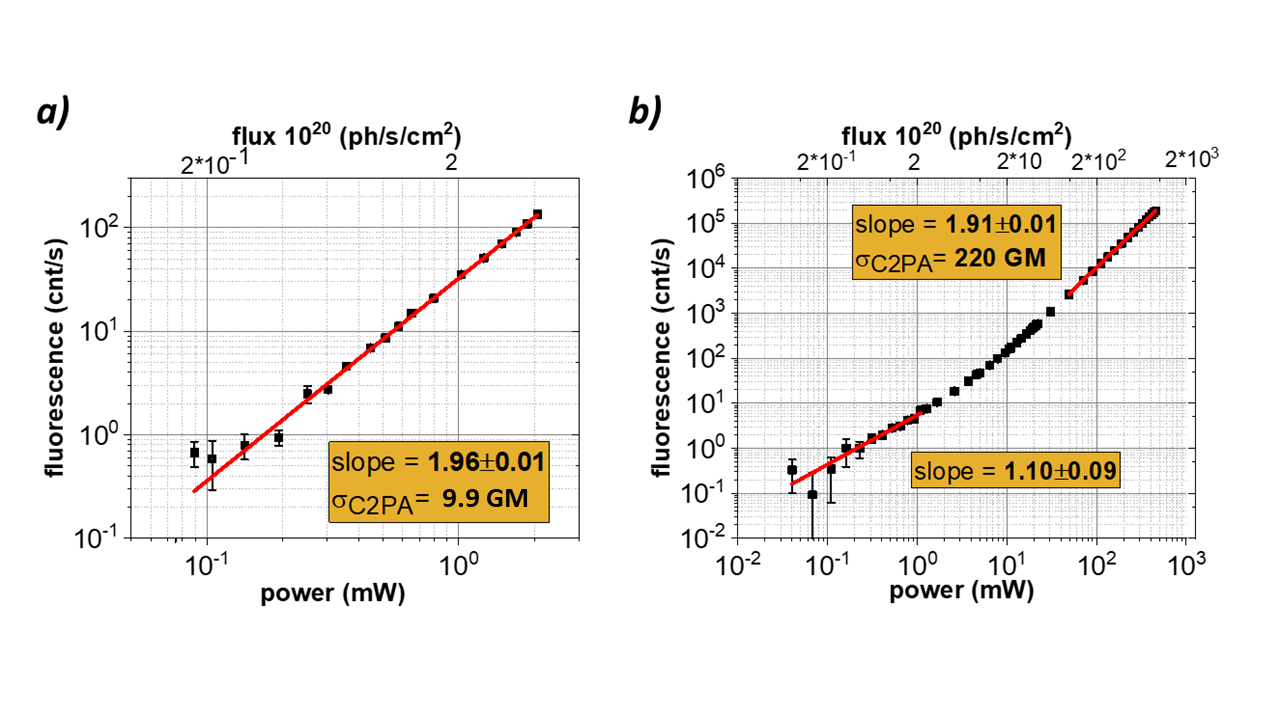}
    \caption{Results of classical (coherent) two-photon excited fluorescence measurements on a log-log scale. Fluorescence signals (vertical axis, in counts per second, cnt/s) versus the laser excitation power (lower horizontal axis, in mW) or versus the excitation photon flux (upper horizontal axis, in photons per second per $\text{cm}^2$) measured for Rh6G and LDS798 are shown in panels \textit{a} and \textit{b}, respectively (black squares). The slope values obtained from the fit (red lines) and the derived cross section values in GM units are indicated in the insets. In the case of LDS798 the slope value changes from quadratic (1.91) to nearly linear (1.10) with decreasing power.}
    \label{fig:C2PEF}
\end{figure}

A detailed description of the experimental setup and the measurement procedures is provided in the Supporting Information (SI) section.
Classical two-photon excited fluorescence (C2PEF) measurements on Rh6G were used to ensure the proper alignment of the optical system and characterize its sensitivity. In Fig.~\ref{fig:C2PEF} the detected fluorescence signal (in counts per second, cnt/s) is plotted versus the excitation photon flux (upper horizontal axis, in photons per second per $\text{cm}^2$) or the excitation power (lower horizontal axis, in mW) on a log-log scale. The minimum count rate that we could assign to fluorescence photons measured above the background level ($3-5$~cnt/s) is determined to be approximately 0.5~cnt/s. The power dependence of C2PEF for Rh6G (Fig.~\ref{fig:C2PEF}~a) in the range of 0.1-2~mW is found to be near-quadratic with a slope (power exponent) of 1.96$\pm$0.01. The sample concentration (1.1~mM), fluorescence quantum yield (0.9~\cite{1987Penzkofer}) and the measured and calculated excitation condition parameters are used to derive the value of the C2PA cross section, $\sigma_{\text{C2PA}} = 9.9$~GM (see details in the SI) which agrees with the literature value of 9.8~GM~\cite{2016deReguardatti}. 

We repeat the C2PEF measurement with LDS798, which has a large $\sigma_{\text{C2PA}}$, but is less advantageous for fluorescence detection because its quantum yield is only 0.054 (see SI) and its emission spectrum is red-shifted from the peak of the detector sensitivity (Fig.~S3). Makarov~$\latin{et~ al.}$~ \cite{2011Makarov} reported $\sigma_{\text{C2PA}}$ of 515~GM for LDS798 excited at 1060~nm. This value was probably overestimated by a factor of two due to an issue with a Rhodamine B reference standard used in that work as was discussed in de Reguardati \textit{et al.}~\cite{2016deReguardatti} 
In our experiment the C2PEF power dependence in the range 50-500~mW is found to have a slope value of 1.91$\pm$0.01 (Fig.~\ref{fig:C2PEF}~\textit{b}). Using a sample of $0.1$~mM LDS798 we derive $\sigma_{\text{C2PA}}=220$~GM.

The methods and apparatus used here are very similar to ones employed in our earlier study~\cite{2021Parzuchowski}, where the uncertainty for determining $\sigma_{\text{C2PA}}$ was estimated to be approximately 28\%. We therefore assume it is similar in the present experiment.

With decreasing excitation power on LDS798, we observe that the slope of the power dependence decreases and reaches a value of 1.10$\pm$0.09 in the 0.05-1~mW range. Overall, the data show a transition from a quadratic (i.e. C2PA) to a linear (i.e. 1PA) excitation regime. Although a transition of this type is rather uncommon in C2PEF experiments, there are several reports of similar behavior indicating the presence of the HBA process~\cite{1983Apatin, 1995Okamura,2003Drobizhev,2007Makarov,2007Rebane,2008Starkey}. As suggested by Drobizhev $\latin{et~  al.}$~\cite{2003Drobizhev} the collected fluorescence signal, $F$ (in cnt/s), can be written as a sum of two terms, one describing the excitation via HBA and the other via C2PA,

\begin{equation}
    F = N K \sigma_{\text{HBA}} \phi + \frac{1}{2} N K \sigma_{\text{C2PA}} \phi^2 
    \label{EqHBAplus2PA}
\end{equation}
where $N$ is the number of molecules in the excitation volume, $K$ is the overall fluorescence collection efficiency, $\phi$ is the excitation photon flux and $\sigma_{\text{HBA}}$ is the HBA cross section. The $\sigma_{\text{HBA}}$ is a function of the excitation frequency $\nu$ and sample temperature $T$ (see the SI for details). Lowering the temperature is expected to decrease the rate of HBA but not affect the rate of C2PA. In our temperature-dependent experiments (see SI) we observe a maximum 12~nm red shift in the steady-state emission spectrum of the fluorophore and 23\% decrease in quantum yield while increasing the temperature, both of which are accounted for in the analysis.

Eq.~\ref{EqHBAplus2PA} indicates that the relative contributions of C2PA and HBA vary with excitation flux. The HBA term depends linearly on excitation power while the C2PA term depends quadratically, thus at higher powers the latter should be dominant. This is consistent with what we observe in our experiment with LDS798 (see more on this below and in SI). 

\begin{figure}[h!]
    \centering
    \includegraphics[width=1\textwidth]{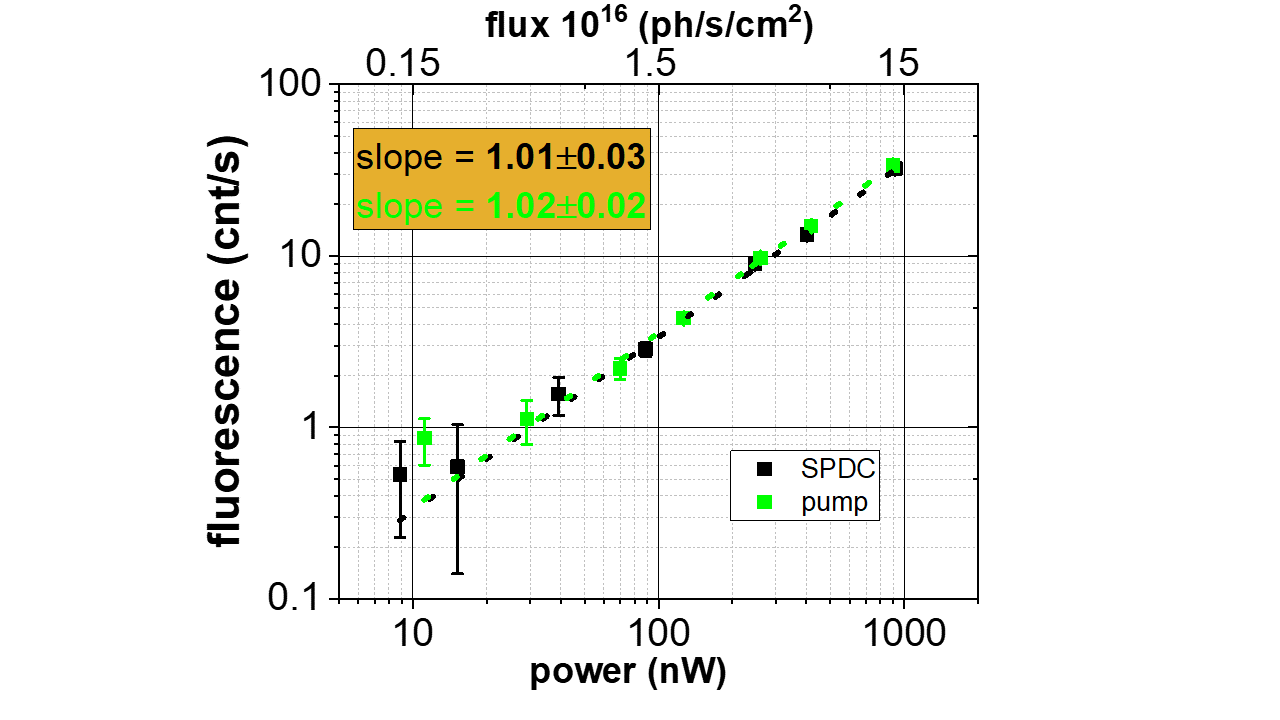}
    \caption{Results of measurements with SPDC excitation of LDS798 on a log-log scale. Fluorescence count rate (in counts per second, cnt/s) versus the SPDC power (lower horizontal axis, in nW) or the excitation photon flux (upper horizontal axis, in photons per second per $\text{cm}^2$). The SPDC excitation flux is calculated assuming the effective wavelength of 1064~nm. The excitation power was controlled by attenuating the SPDC (black squares) or pump laser (green squares) beams. In both cases the signal dependence was linear as determined by the slope values (shown in the inset) calculated from the fits (dashed black and green line, respectively). }
    \label{fig:HBAwithSPDC}
\end{figure}

Next, we block the 1060~nm laser and switch to SPDC excitation. For Rh6G we are unable to detect any fluorescence signal with the maximum available SPDC power of approximately 1.3~$\mu$W. 
However, for LDS798  we measure a strong fluorescence signal (up to 40~cnt/s) under SPDC excitation (Fig.~\ref{fig:HBAwithSPDC}). 
First, we carefully verify that this signal is not a scattered portion of the SPDC light and not related to the solvent itself. Replacing the LDS798 sample with pure $\text{CDCl}_{3}$ results in no signal observed above the background level.
To assess whether the signal from the LDS798 sample is E2PEF, we test for a unique signature of the process as discussed above. In two separate measurements, we vary the SPDC pump power and attenuate the SPDC beam. Figure~\ref{fig:HBAwithSPDC} is a plot of the measured fluorescence signals (vertical axis, in cnt/s) versus the excitation power (lower horizontal axis, in nW) or, equivalently, versus the flux (upper horizontal axis, in photons per sec per $\text{cm}^2$) on a log-log scale. Varying the SPDC pump power (green squares), we observe that the fluorescence follows a linear dependence (slope value of 1.02$\pm$0.02), which is consistent with E2PEF. However, attenuation of the SPDC beam power (black symbols) also results in a linear dependence (slope value of 1.01$\pm$0.03). The latter result clearly indicates that the fluorescence signal is not related to E2PA because attenuating the SPDC beam with a neutral density filter randomly removes individual photons rather than photon pairs, thus making the excitation more classical, and classical 2PA scales quadratically with power.

We perform two additional experiments that confirm the observation of HBA. We characterized the temperature-dependence of the C2PA signal on LDS798 encapsulated in a poly-dimethylsiloxane (PDMS) matrix. A rigid polymer rather than a liquid is selected to ensure that changes in solvent viscosity with temperature do not influence the radiationless relaxation rate of LDS798 and thus its fluorescence quantum yield.~\cite{Doan2017} The fluorescence signal in PDMS was observed to increase nearly four-fold with increasing temperature from 283~K to 323~K, and is well fit by a Boltzmann function (Fig.~S4~b (SI)). The experiment was repeated with SPDC excitation (Fig.~S4~c (SI)). The measured signal scales with temperature in the same manner. In addition, we used an independent setup designed for characterizing absorption cross sections~\cite{2020Drobizhev} to measure the HBA cross section of LDS798 as a function of wavelength in the 680 to 900~nm region. We compared this to the cross section we derived from the data shown in Fig.~\ref{fig:C2PEF}~b. The cross sections in the red tail region, including our 1060~nm data point, fit to a Boltzmann function (Fig.~S8 (SI)), which is consistent with HBA theory (Eq.~6 (SI)). Finally, we note that a model entirely based on HBA without any adjustable parameters is consistent with both the laser-excited and SPDC-excited fluorescence signals (Fig.~S7 (SI)). 

Several important points can be concluded from this study. 
We have shown that even when the excitation wavelengths are detuned hundreds of nanometers from the 1PA peaks of a chromophore, vibronic states can still be excited via HBA. Although this effect is known from previous reports on C2PA, it has not been discussed in previous studies of E2PA. Explaining the origin of inconsistency among different experiments is the most significant challenge currently facing the development of E2PA spectroscopy and its applications. As shown here for LDS798, the HBA signal can partially mimic the power scaling of E2PA. It seems likely that this mechanism could be contributing to E2PA measurements on any other chromophore. Potential HBA contributions should be carefully quantified since they could lead to a significant over-estimate of the quantum enhancement for the 2PA efficiency. Our results underline a critical need to perform stringent tests for unique signatures of E2PA in measured signals with SPDC excitation to distinguish one-photon processes from E2PA. 
In particular, to confirm a signal is from E2PA as opposed to other potential mechanisms, the proper validation procedure is to vary the incident power from the entangled photon source both by attenuating the power input to the SPDC crystal and also by attenuating the power afterwards. To demonstrate E2PEF, these two methods of varying the incident power must show different fluorescence power dependencies.

\section{Supporting Information}
Details of the experimental setup, applied measurement procedures, $\sigma_{\text{C2PA}}$ calculations, temperature dependence measurements, modeling the measured HBA signal and measurements of HBA cross section as a function of wavelength are available in the supporting information section.
\begin{acknowledgement}
We acknowledge Mikhail Drobizhev (Montana State University) for suggesting that we closely examine the HBA contributions and for the technical help with some of the measurements. Some of the experimental results were obtained using his resource for multiphoton characterization of genetically encoded probes, supported by the NIH/NINDS grant U24NS109107.
AM and KMP thank Srijit Mukherjee for assisting with the 1PA measurements and for valuable discussions in preparation of the manuscript. This work was supported by NIST and by the NSF Physics Frontier Center at JILA (PHY 1734006) and by the NSF-STROBE center (DMR 1548924). 
 
Certain commercial equipment, instruments, or materials are identified in this paper in order to specify the experimental procedure adequately. Such identification is not intended to imply recommendation or endorsement by NIST, nor is it intended to imply that the materials or equipment identified are necessarily the best available for the purpose.

\end{acknowledgement}

\section{Present Addresses}

A.M.: Max Planck Institute for the Science of Light, Staudtstrasse 2, 91058 Erlangen, Germany.

\beginsupplement

\begin{suppinfo}
%--------------------------------------

\section{Experimental setup and measurement methods}

\begin{figure}
    \centering
    \includegraphics[width=1\textwidth]{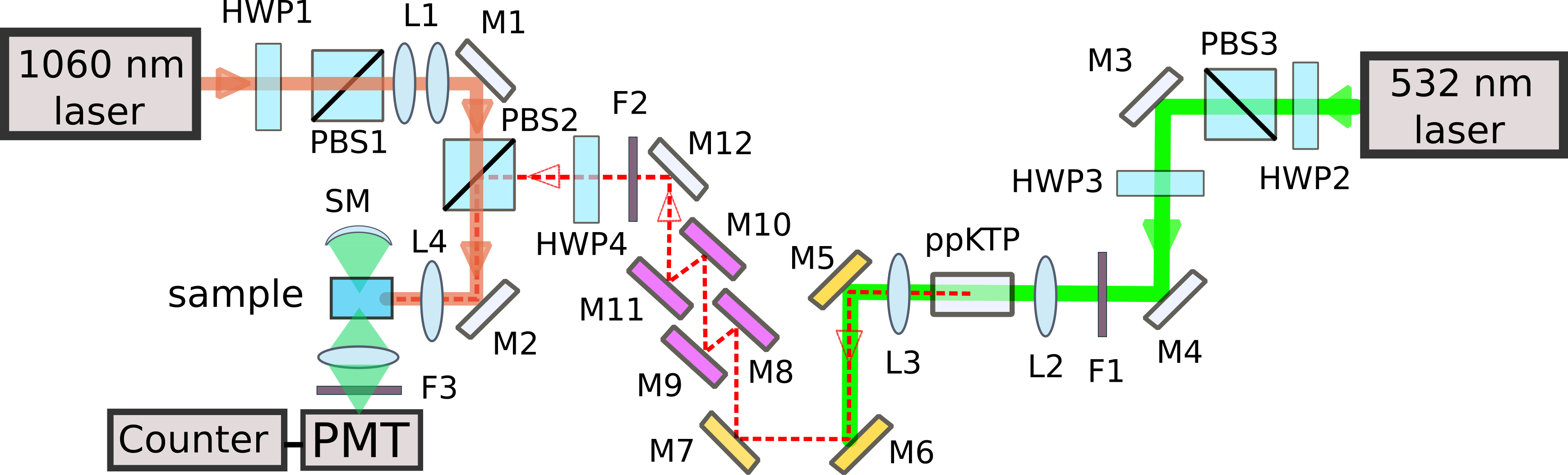}
    \caption{Schematic of the experiment setup. The 1060~nm laser, SPDC pump and SPDC beams are indicated by solid red, solid green and dashed red lines respectively. See text for details.}
    \label{fig:setup}
\end{figure}

As schematically shown in Fig.~\ref{fig:setup}, the experimental setup includes two excitation sources. The fiber laser provides CW 1060~nm radiation (solid red line) for the C2PEF measurements. Its power is controlled with a half-wave plate (HWP1) placed inside a motorized rotational stage and a Glan-Thompson polarizer cube (PBS1). Back reflections from the power controlling optics are blocked from re-entering the fiber output with a Faraday isolator (not shown). Two lenses forming a telescope (L1) are used to adjust the laser beam size. At the polarizing cube (PBS2) the laser beam is overlapped with the SPDC beam (dashed red line). 

A CW pump laser at 532~nm (solid green line) with a maximum output of 5W is used as the pump. The laser power is controlled with a half-wave plate (HWP2) and a polarizing cube (PBS3). A band pass filter (F1), and 3 dichroic mirrors are added to remove harmonics generated by the strong pump beam anywhere in the optical system. The pump polarization is controlled with a half-wave plate (HWP3) before entering a 12~mm periodically poled potassium titanyl phosphate (ppKTP) crystal designed for Type-0 SPDC. The pump laser beam is focused with a lens (L2) to full width at half maximum ($\text{FWHM}$ ) $\approx30$~$\mu$m inside the crystal. The ppKTP crystal is placed inside a custom made holder, and its temperature is stabilized at 25~$^{\circ}\mathrm{C}$ with a recirculating chiller. The pump polarization and the crystal temperature are adjusted to optimize the SPDC generation, see Fig.~\ref{fig:SPDCspectrum}~a. The generated SPDC light has a spectral FWHM of 128.9~nm centered at 1077.4~nm (Fig.~\ref{fig:SPDCspectrum}~b). The SPDC spectrum is measured with a InGaAs spectrometer used with a fiber-coupled input. The SPDC light is collimated with a lens (L3). The remaining pump is filtered out with a series of dichroic mirrors (M5--M7) and several long pass interference filters (F2). The total OD at 532~nm is greater than 27. The accumulated group velocity dispersion (GVD) for the SPDC beam that traveled through the setup to the sample position is approximately 1600~$\text{fs}^2$ at the central wavelength. This GVD is compensated via multiple reflections between four chirped mirrors (M8--M11). The GVD is fine-tuned with a 4~mm-thick YAG window. We estimate the short- and long-wavelength edges of the SPDC spectrum have $\approx 150$~fs$^2$ of uncompensated GVD. The polarization of the SPDC is controlled with a half-wave plate (HWP4) before PBS2. The maximum pump power corresponds to 1.3~$\mu$W of output SPDC power at the sample, a value similar to previous E2PA reports~\cite{2021Tabakaev,2021Landes}. While characterizing the entanglement and the photon pair production rate of this source are beyond the scope of this study, the linear losses in the system provide an estimated 17.6\% transmission efficiency for each SPDC photon by the time it reaches the sample resulting in an estimated 40~nW of SPDC power (3.15\% transmission for pairs).

The 1060~nm laser and SPDC beams are focused with a lens (L4) into the sample. At the focus inside the sample cuvette, the laser and SPDC beam widths are 55.5~$\mu$m and 67~$\mu$m FWHM, respectively. Several silver mirrors (M1--M4 and M12) are used to steer the beams to the sample position. The sample is contained in a 2$\times$10~mm spectroscopic quartz cuvette. Fluorescence is collected perpendicular to the beam propagation direction, in a similar manner to that detailed in Parzuchowski \textit{et al.}~\cite{2021Parzuchowski} A spherical mirror (SM) is used to increase the fluorescence collection. From ray-tracing simulations using the Zemax OpticStudio software, we estimate that the geometrical fluorescence collection efficiency is approximately 4.2$\%$. The fluorescence is detected with a photon counting photomultiplier tube (PMT) cooled to 5~$^{\circ}\mathrm{C}$ using a thermoelectric chiller. Details regarding filters placed in front of the PMT (F3), the sample emission, and the PMT quantum efficiency are provided in Fig.~\ref{fig:overplap}. The digitized counts are measured with a counter. The laser and the SPDC powers are measured with a silicon photodiode power sensor calibrated relative to a germanium photodiode. The SPDC beam is attenuated with ND filters introduced right after the F2 filters. The PMT background counts are measured by blocking the excitation beams before the sample. The fluorescence and the background counts at each power are averaged over 100~seconds total. For the regime when the fluorescence counts are 10 cnt/s or lower this is increased to 1000~sec to reduce the standard deviations. The measurements and the data acquisition are controlled with LabVIEW.
 \begin{figure}
 \centering
 \includegraphics[width=1\textwidth]{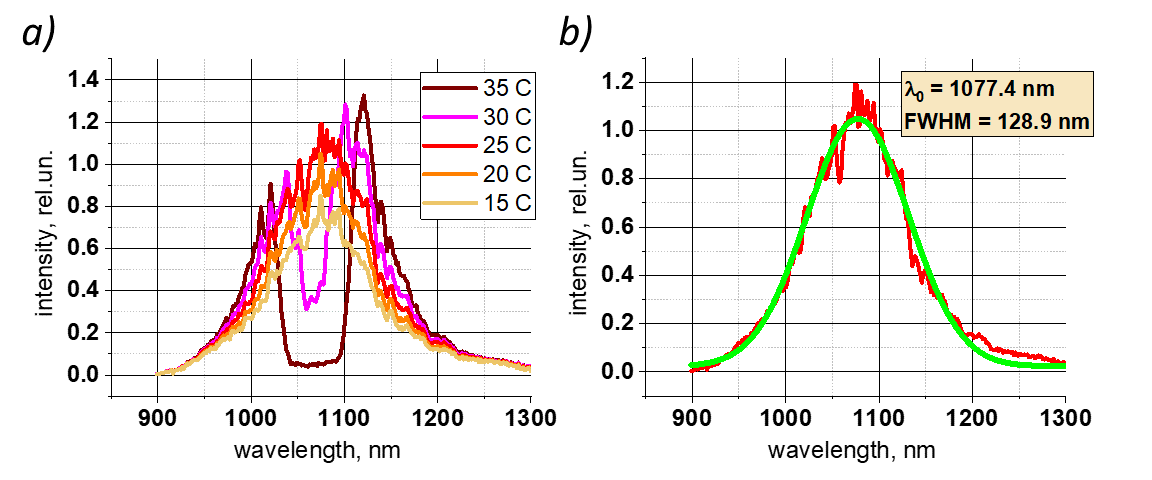}
 \caption{Measured SPDC light spectrum. The spectrum as a function of the ppKTP crystal temperature is shown in panel a. The chosen temperatures are shown in the legend. At 25~$^{\circ}\mathrm{C}$ the spectrum is symmetric with the central wavelength $\approx1077.4$~nm and FWHM of 128.9~nm (panel b) determined from a Gaussian fit (solid green curve). }
 \label{fig:SPDCspectrum}
\end{figure}

Samples of Rh6G (SigmaAldrich) and LDS798 (Luxottica) were prepared in methanol (Thermo Fisher Scientific, ACS grade) and $\text{CDCl}_{3}$ (Cambridge Isotope Laboratories, 99.8$\%$) respectively. 
Rh6G and LDS798 concentrations were 1.1~mM and 0.3~mM, respectively.
Chromophore concentrations were determined spectrophotometrically. We observed about 10\% maximum change in the concentration over the course of the measurements due to solvent evaporation or photobleaching. For the sample concentrations used in our study the linear absorption spectra closely agreed with available literature data. For Rh6G, a previous study found that formation of aggregates at this concentration is negligible.~\cite{1987Penzkofer} We found no evidence of aggregate formation for these sample concentrations. 

%------------------------------------------------------------
\section{Rh6G and LDS798 emission, PMT quantum efficiency and transmission spectra of optical filters}
Here we provide the emission spectra of the samples, PMT quantum efficiency and transmission spectra of the filters. These spectra are shown in a single plot for each of the samples in Fig.~\ref{fig:overplap}. The SPDC spectrum (dashed magenta) and the samples' absorption (dashed red) and emission (dashed blue) spectra are not shown to scale (relative shapes). The PMT sensitivity curve (black) and various used Semrock filters (see legends for the color mapping) are also shown. The left vertical axis indicates the PMT's quantum efficiency (in \%) whereas the right vertical axis shows optical density, OD, of the filters used in the experiment. Note for panel b: a FF01-795/188 filter was originally used for the C2PEF measurements of LDS798 and was later replaced with a FF01-709/167 for the E2PEF study.

\begin{figure}
     \centering
     \includegraphics[width=1\textwidth]{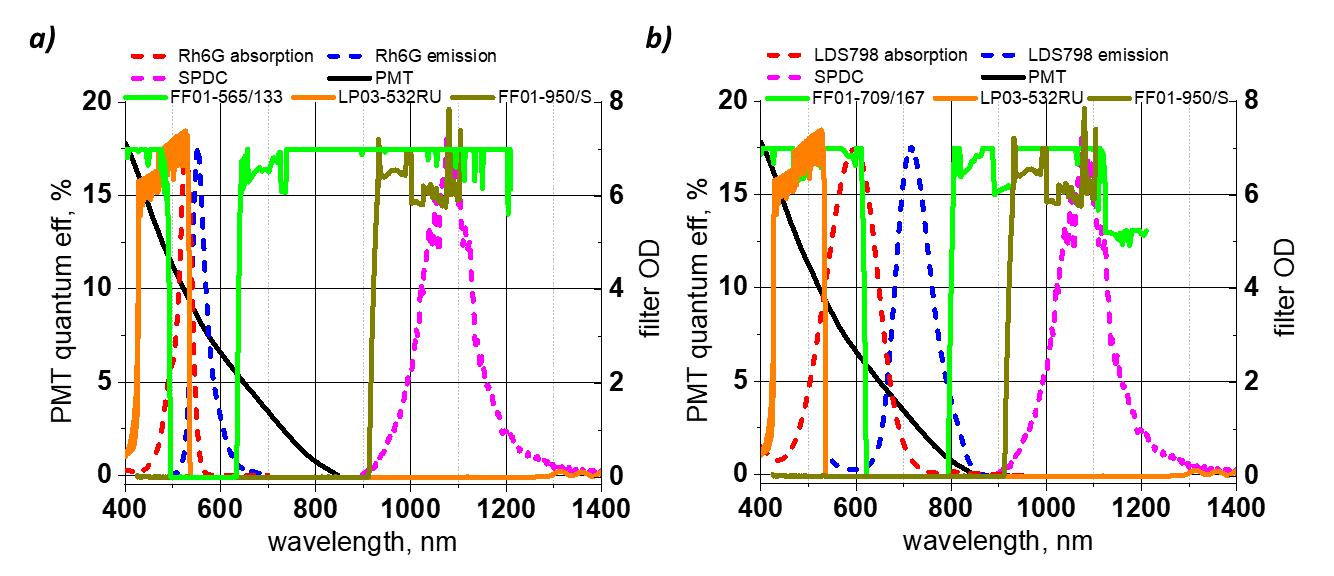}
     \caption{Overlap of the samples' absorption (dashed red) and emission spectra (dashed blue), filter spectra (green, orange, brown), PMT quantum efficiency (black) and SPDC spectrum (dashed magenta) for Rh6G (a) and LDS798 (b) experiments.}
     \label{fig:overplap}
 \end{figure}

%--------------------------------------------------
\section{Calculation of C2PA cross sections}
Here we detail how $\sigma_{\text{C2PA}}$ for Rh6G and LDS798 are calculated using the experimental data. Assuming a Gaussian spatial profile for the excitation CW laser beam we can express the collected fluorescence signal (in cnt/s) excited via C2PA as \cite{2011Makarov,2021Parzuchowski}
\begin{equation}
\label{SIF2}
    F_2= 2^{1/2}\times \left(\frac{\log{2}}{\pi}\right)^{3/2} \times \gamma \times \kappa  \times \eta \times N_{\text{mol}} \times L  \times \sigma_{\text{C2PA}} \times \frac{1}{S} \times {N_\text{ph}}^2 
\end{equation}
where 
$\kappa$ and $\gamma$ (both dimensionless) are geometrical (setup specific) and optical (setup and sample specific) fluorescence collection efficiencies, respectively; the latter is a function of the emission frequency $\nu_{\text{em}}$; the product of these parameters is the absolute fluorescence collection efficiency (shows what portion of the emitted photons is registered); $\eta$ is the sample's quantum yield (dimensionless); $N_{\text{mol}}$ (in $\text{m}^{-3}$) is the number density of molecules and $L$ is the sample's thickness; $S$ is the beam area and $N_{\text{ph}}$ is the number of excitation photons incident on the sample per second. For convenience $N_\text{mol}$ can be re-written in terms of molar concentration $C_m$ (in mol $L^{-1}$) and Avogadro's number $N_A$ as $N_\text{mol}=N_A \times C_m \times 10^3$. For the CW excitation regime, $N_{\text{ph}}$ is simply excitation power $P$ (in W) per excitation photon energy $h \nu$ (in J). The values of the experimental parameters are measured or calculated, and the cross section can be calculated from here as
\begin{equation}
    \sigma_{\text{C2PA}}= \frac{F_2}{2^{1/2}\times \left(\frac{\log{2}}{\pi}\right)^{3/2} \times \gamma \times \kappa \times \eta \times N_A \times C_m \times 10^3 \times L \times \frac{1}{S}\times \left(\frac{P}{h \nu}\right)^2}
\end{equation}

Now we can insert numbers for our experiment.
The beam size measured at the sample position in vertical and horizontal projections have FWHM of 55 and 57~$\mu$m. We approximate the beam area $S$ (in $\text{m}^2$) as 
\begin{equation}
    S\simeq \pi \times \frac{55}{2} \times \frac{57}{2} \times 10^{-12} 
\end{equation} 
The Rayleigh range of the laser was measured to be 6400 $\mu$m, thus the beam area changes by less than a factor of 2 within the 1~cm cuvette. We obtain the geometrical collection efficiency $\kappa$ using a Zemax simulation similar to those detailed in Parzuchowski \textit{et al.}~\cite{2021Parzuchowski}. In this simulation we assume that the beam is of uniform area (55~$\mu$m FWHM) throughout the cuvette and arrive at $\kappa \approx$~0.042. We did not measure $\kappa$, and may expect a similar deviation of the experimental and simulated values as that found in Parzuchowski \textit{et al.}, where the experimental $\kappa$ was $\approx 23 \%$ lower than the simulated value.

Next we proceed with the sample specific parameters. 
First, consider Rh6G.
The value of $\gamma$ is found from the multiplication of the normalized emission profile of Rh6G, absolute transmission curves for used fluorescence filters and the PMT's quantum efficiency at the corresponding wavelength (Fig.~\ref{fig:overplap}). This gives $\gamma \approx 0.075$. The quantum yield $\eta$ value is $\approx 0.9$~\cite{1987Penzkofer}. For the measurement we used a Rh6G sample of concentration $C_m \approx 1.1$~mM.
To calculate an averaged value of $\sigma_{\text{C2PA}}$ we fit the fluorescence signals measured over the range of 0.1-2~mW shown in the Fig.~2~a (main text). From here we estimate $\sigma_{\text{C2PA}} \approx 9.9$~GM.

Next we repeat the same for the LDS798 sample. In this case we determine $\gamma \approx 0.018$. The quantum yield for LDS798 in $\text{CDCl}_{3}$ is measured using LDS798 in EtOH as a reference, which has a quantum yield of 0.011~\cite{2009Luchowski}. The determined $\eta$ value is $\approx 0.054$. For the C2PA measurement we used a LDS798 sample of concentration $C_m \approx 0.1$~mM. For averaging we use the quadratic portion of the measured fluorescence signals corresponding to 50-500~mW (Fig.~2~b (main text)). We estimate $\sigma_{\text{C2PA}} \approx 220$~GM.
%%%%%%%%%%%%%%%%%%%%%%%%%%%%%%%%%%%%%%
\section{Sample temperature dependence of HBA}

\begin{figure}
     \centering
     \includegraphics[width=1\textwidth]{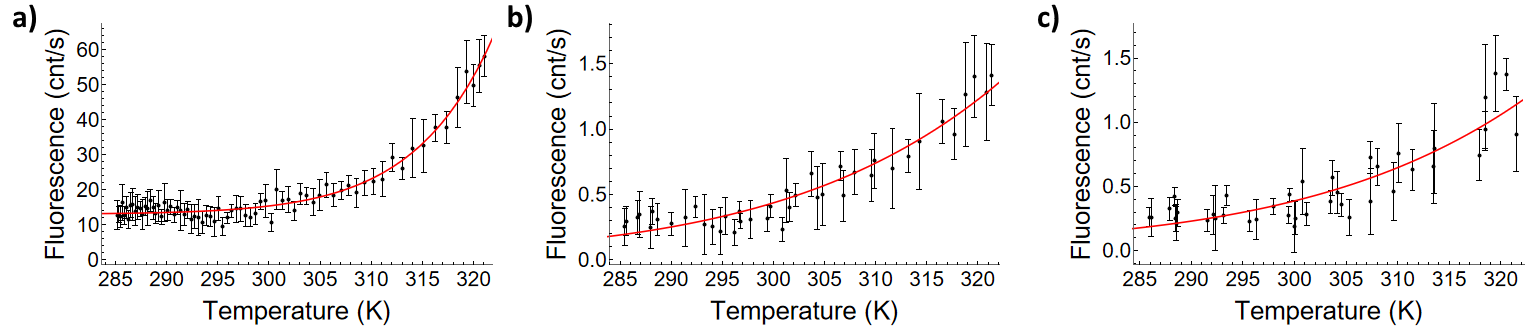}
     \caption{Panel a: fluorescence counts (in cnt/s) versus the sample temperature (in K) obtained using 10~mW of the 1060~nm laser power. The red curve shows the exponential dependence of the fluorescence verifying its HBA origin in this regime. Panel b: the same as the panel a except the laser source is replaced with the SPDC. Panel c: the same as the panel b except a 1055~nm short-pass filter was placed in the SPDC beam path.}
     \label{fig:ExpVsWavel}
 \end{figure}

To further confirm that the fluorescence signal that we measure is HBA, we vary the LDS798 sample temperature. For this experiment LDS798 was encapsulated in a molded  PDMS slab. To prepare this sample, a pristine slab of PDMS was cast to fit the cuvette holder, and then soaked in a 2.3~mM solution of LDS798 in chloroform. The temperature of the solid sample was measured using a thermistor cast directly into the PDMS sample. This allowed for simultaneous measurement of temperature and fluorescence rates. We used the 1060~nm laser at 1~mW of power. This power is low enough that C2PEF contribution should be negligible (see Fig.~2~b (main text)). The temperature was first held at 50~$^{\circ}\mathrm{C}$, then ramped down to 10~$^{\circ}\mathrm{C}$, and finally ramped back up to 50~$^{\circ}\mathrm{C}$ to ensure no sample degradation occurred. 

The population of LDS798 vibronic states is expected to follow Boltzmann statistics. Therefore increasing or decreasing the temperature should result in an increase or decrease of HBA, respectively. The measured fluorescence counts versus temperature for laser excitation (Fig.~\ref{fig:ExpVsWavel}~a) illustrate precisely this behavior. The fit (red curve) for the fluorescence signal $F$ is of the form  $F = A \times \exp{(-E/kT)} + C$, where $E$ is the energy of a 1064~nm photon, $k$ is the Boltzmann constant, and $A$ and $C$ are the fitting parameters. The same experiment is then repeated using SPDC excitation (Fig.~\ref{fig:ExpVsWavel}~b) and SPDC excitation filtered with a 1055~nm short-pass filter (Fig.~\ref{fig:ExpVsWavel}~c). In both cases the signal follows a Boltzmann temperature-dependence. These results indicate the fluorescence orginates from HBA for both laser and SPDC excitation. Alternatively, to confirm that HBA occurs when the sample is excited by SPDC photon pairs, one could measure the fluorescence signal as a function of spectral width at a fixed temperature. In this case, one would expect the signal to decrease as the photon spectrum becomes narrower.

\section{Simulations of HBA signals}
We calculate the fluorescence signals expected from HBA with laser and SPDC excitation. 
The collected one-photon excited fluorescence signal (in cnt/s), using excitation photons of frequency $\nu$ (in Hz) spread out over the range of $\Delta\nu$ (in Hz), can be written as

\begin{equation}
\label{SIF1}
F_1=\gamma \times \kappa \times \eta \times N_{\text{mol}} \times L\times \int\displaylimits_{\Delta \nu} \sigma_1 (\nu)  \times  \mathcal{F}(\nu) d\nu
\end{equation}
where $\sigma_1$ is the 1PA cross section (in $\text{cm}^2$), $\mathcal{F}(\nu)$ is the flux spectral density (in photons per Hz per s) and the remaining notation is the same as in Eq.~\ref{SIF2}. For the CW case

\begin{equation}
    \mathcal{F} (\nu)=\frac{\mathcal{P} (\nu)}{h\nu}
\end{equation}
where $\mathcal{P}(\nu)$ is a power spectral density (in W per Hz) which can be calculated from the measured total power, $P$, (in W) and a dimensionless spectral power density function, $f(\nu)$, obtained from the simulated SPDC spectrum (Fig.~\ref{fig:mirroredSpectrumHBA}~a). 

For HBA the value of $\sigma_1$ can be written as~\cite{2003Drobizhev,2005Kachynski}
\begin{equation}
\label{SIsigmaHBA}
    \sigma_1=\sigma_{\text{max}} \times \exp{\left[ \frac{-h (\nu_{\text{max}}-\nu)}{kT}\right]} \times \text{FC}(\nu)
\end{equation}
where $\sigma_{\text{max}}$ is $\sigma_1$ at the frequency of the ``0--0" transition $\nu_{\text{max}}$; $k$ is Boltzmann's constant and $T$ is the sample's temperature. FC is a normalized Franck-Condon factor (see below). The value of  $\sigma_{\text{max}}$ (in $\text{cm}^2$) is calculated from the extinction coefficient $\epsilon_{\text{max}}$ (in $\text{M}^{-1} \text{cm}^{-1}$) at $\nu_{\text{max}}$ as $\sigma_{\text{max}}= \epsilon_{\text{max}} \times 3.82 \times 10^{-21}$~\cite{2006Lakowicz}.

\begin{figure}
     \centering
     \includegraphics[width=1\textwidth]{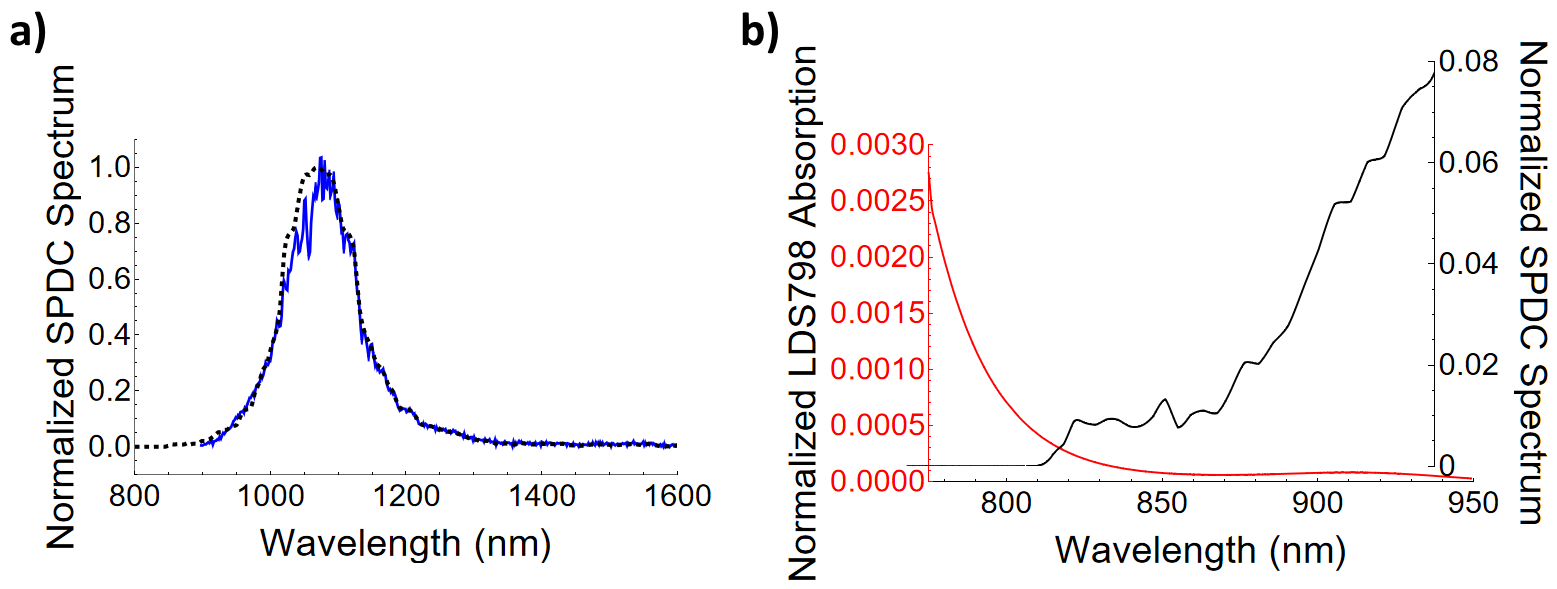}
     \caption{Panel a: The measured SPDC spectrum (blue dashed) overlaid with the mirrored spectrum (black solid) as a function of wavelength. The mirrored spectrum is constructed by reflecting the red side of the spectrum about the central SPDC wavelength in order to obtain an approximation for the tail on the blue side. Panel b: The reflected portion of the SPDC spectrum (black) overlaid on the red tail of the LDS798 absorption spectrum. The reflected spectrum is cut off around 850~nm due to a long-pass filter in place before the sample.}
     \label{fig:mirroredSpectrumHBA}
 \end{figure}

Here $\gamma \approx0.082$ for Rh6G and $\gamma \approx 0.025$ for LDS798, which is slightly larger for this measurement because a different filter was employed compared to the C2PEF measurements described above (see Fig.~\ref{fig:overplap}~b). 
The beam size of the SPDC beam is measured to be 65~$\mu$m FWHM at its focus in the sample, and the Rayleigh range is 920~$\mu$m. This beam is significantly more divergent than the laser beam, leading to a different shape of the excitation volume within the cuvette. Since the collection optics collect fluorescence most efficiently at the center of the cuvette (as shown in Parzuchowski \textit{et al.}~\cite{2021Parzuchowski}) where the excitation volumes look much more similar, we assume that $\kappa \approx 0.042$, which was simulated for laser excitation, can be used here as well.
For Rh6G $\eta \approx 0.9$ and was measured using a sample concentration of $C_m \approx 1.1$~mM and a path length of 1~cm. The simulated HBA fluorescence signal from 1~mW of classical excitation is $8.9\times 10^{-3}$~cnt/s, which is consistent with the lack of linear dependence in Fig.~2~a. Using equation 4, we estimated that the signal due to HBA with 1.3~$\mu$W SPDC power is 2 orders of magnitude below background level. Therefore, 130~$\mu$W would be necessary to generate a measurable signal with SPDC excitation. Based on our calculations, a sample temperature above $80^{\circ}$C would be necessary to observe a linear regime with classical excitation or similarly to observe a signal above background with SPDC excitation.
For LDS798 $\eta \approx$~0.054. For the measurement we used a LDS798 sample of concentration $C_m \approx 0.3$~mM. As before we approximate the value $L$ to be equal to the cuvette length (1~cm).

The FC($\nu$) value can be estimated from the ratio of the sample absorption spectrum at $2\nu_{\text{max}}-\nu$ and $\nu_{\text{max}}$~\cite{2003Drobizhev}. The value of $\nu_{\text{max}}$ can be determined from the intersection of the properly normalized emission and absorption profiles~\cite{2006Lakowicz} and is found to correspond to approximately 672~nm, Fig.~\ref{fig:FrankCondon}~a. The corresponding extinction coefficient is $\epsilon_{\text{max}} \approx 1.54\times 10^4$~$\text{M}^{-1}\times~\text{cm}^{-1}$. 

\begin{figure}
     \centering
     \includegraphics[width=1\textwidth]{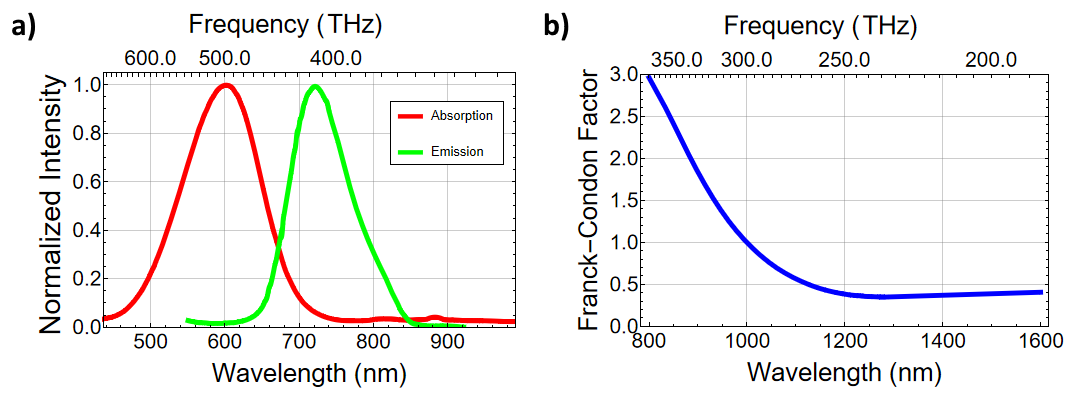}
     \caption{Determination of the Franck-Condon parameters ratio (FC). The intersection of the normalized emission and absorption spectral profiles of LDS798 is used to find the $\nu_{\text{max}}$ position (a) and then to plot the FC value as a function of frequency $\nu$ (in sec$^{-1}$) according to Drobizhev \textit{et al.}~\cite{2003Drobizhev} (b).}
     \label{fig:FrankCondon}
 \end{figure}

For LDS798 we estimate the FC value falls in the range between 0.41 and 2.65 in the $\Delta \nu \approx (200-361.4)$~$\text{THz}$ range (Fig.~\ref{fig:FrankCondon}~b).

When calculating the expected fluorescence from HBA excited by SPDC, a numerical integration has to be done due to the wide bandwidth which results in significantly different contributions from the red and blue sides of the spectrum. The calculated fluorescence signal is found to be particularly sensitive to the low amplitude tail of the SPDC spectrum on the blue side of the spectrum which is difficult to measure accurately due to the limited sensitivity of the spectrometer (based on a thermoelectrically-cooled linear InGaAs array) in this spectral region. To account for this spectral content, we employ a model SPDC spectrum in which the red side of the measured spectrum is reflected about the central frequency. The spectrum of this type of degenerate SPDC source is theoretically predicted to be symmetric~\cite{2013Lerch,2021Szoke}. Then the fluorescence signal $F_{\text{HBA}}$ is given by 

\begin{equation}
\label{SIF1tot}
\begin{split}
    F_{\text{HBA}} = \gamma \times \kappa \times \eta \times C_m \times 10^3 \times L \times \epsilon_{\text{max}} \times \ln 10 \times \hspace{10em} \\
    \qquad \times \int\displaylimits_{\Delta \nu} 
    \left( \exp{\left[ \frac{-h (\nu_{\text{max}}-\nu)}{kT}\right]} \times \text{FC}(\nu) \right)  \times \mathcal{F}(\nu) ~d\nu
\end{split}
\end{equation}

As already mentioned, the SPDC spectral density, $\mathcal{F}(\nu)$  (in photons/sec/nm), was obtained by measuring the red side of the spectrum and reflecting it about the central wavelength (Fig.~\ref{fig:mirroredSpectrumHBA}~a). The region from 1550~nm to 1600~nm was used to ensure proper background subtraction of the SPDC spectrum which corresponds to a spectral cutoff at 850~nm on the reflected blue side. The calculated fluorescence is consistent with measured values for SPDC induced HBA (Fig.~3 (main text)) being nearly 24-fold lower (Fig.~\ref{fig:MeasuredVsCalcHBA}).

\begin{figure}
     \centering
     \includegraphics[width=1\textwidth]{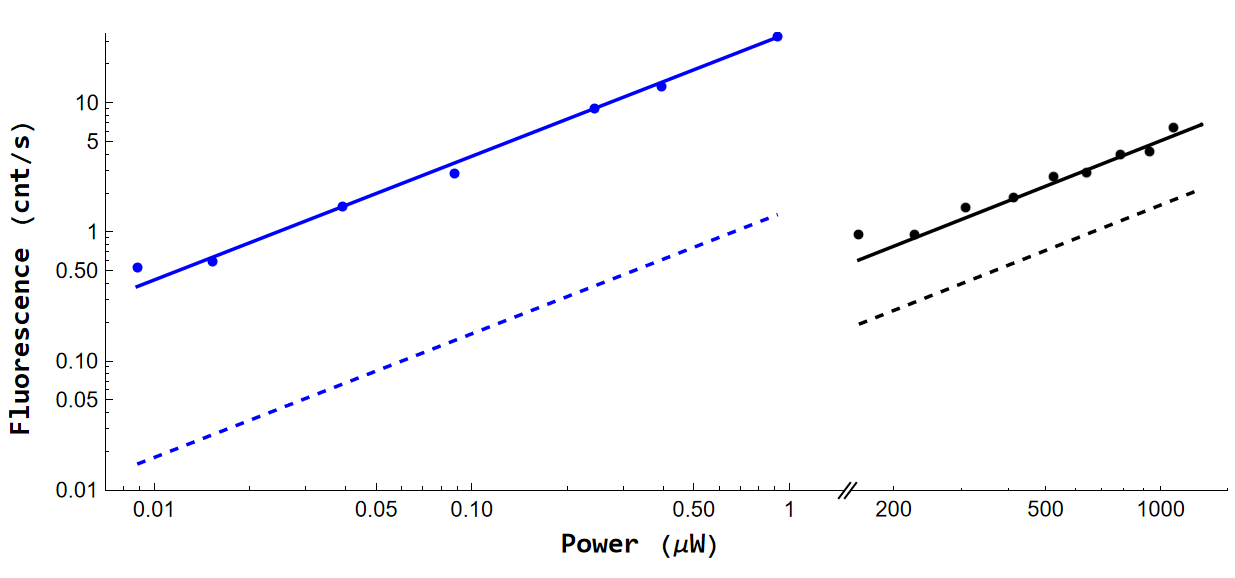}
     \caption{Calculated (dashed) and measured (with solid fit line) fluorescence counts (in cnt/s) versus laser power (in~$\mu$W) for both the 1060nm laser (black) and SPDC source (blue). The calculated fluorescence signal is 24-fold lower than the experimental value for the SPDC source and 3-fold lower for the classical source.}
     \label{fig:MeasuredVsCalcHBA}
 \end{figure}
The calculated HBA cross section is subject to uncertainty because the method used to determine $\nu_{\text{max}}$ is based on the mirror-image rule~\cite{2006Lakowicz}, and without investigating the electronic structure of LDS798 it is not clear if this holds.
At the same time, the method used to model the blue tail of the SPDC spectrum relies on the far-red tail, a spectral region which was lossy in our setup due to filters used to block the pump beam.
Because the calculation is highly sensitive to changes in the values of these parameters, it is believed that these issues are the main cause of discrepancy between measured and calculated rates. For example, a 15~nm blue shift on the long-pass filter used would be enough to increase the calculated signal an order of magnitude bringing it within 3-fold of the measured value.
It was also assumed that the collection efficiency for both the classical and SPDC excited fluorescence were identical, though in reality this is most likely not the case due to small mismatches in beam shape and divergence.

Next, the same calculation is repeated to simulate HBA with the 1060~nm laser source assuming a Gaussian spectrum with a 1~nm FWHM.
For consistency, the same spectral integration is used for the classical laser, though is unlikely to be necessary due to the narrow bandwidth.
This resulted in a calculated fluorescence signal that was 3-fold lower than the measured value (Fig.~\ref{fig:MeasuredVsCalcHBA}).

\section{Measured HBA cross section}
We compared the derived HBA cross section for LDS798 from our CW laser-based experiment (Fig.~2~b (main text)) to those obtained over a range of wavelengths using an independent experimental setup. These cross sections are plotted in Fig.~\ref{fig:1PAcrossSection}.

\begin{figure}
     \centering
     \includegraphics[width=0.7\textwidth]{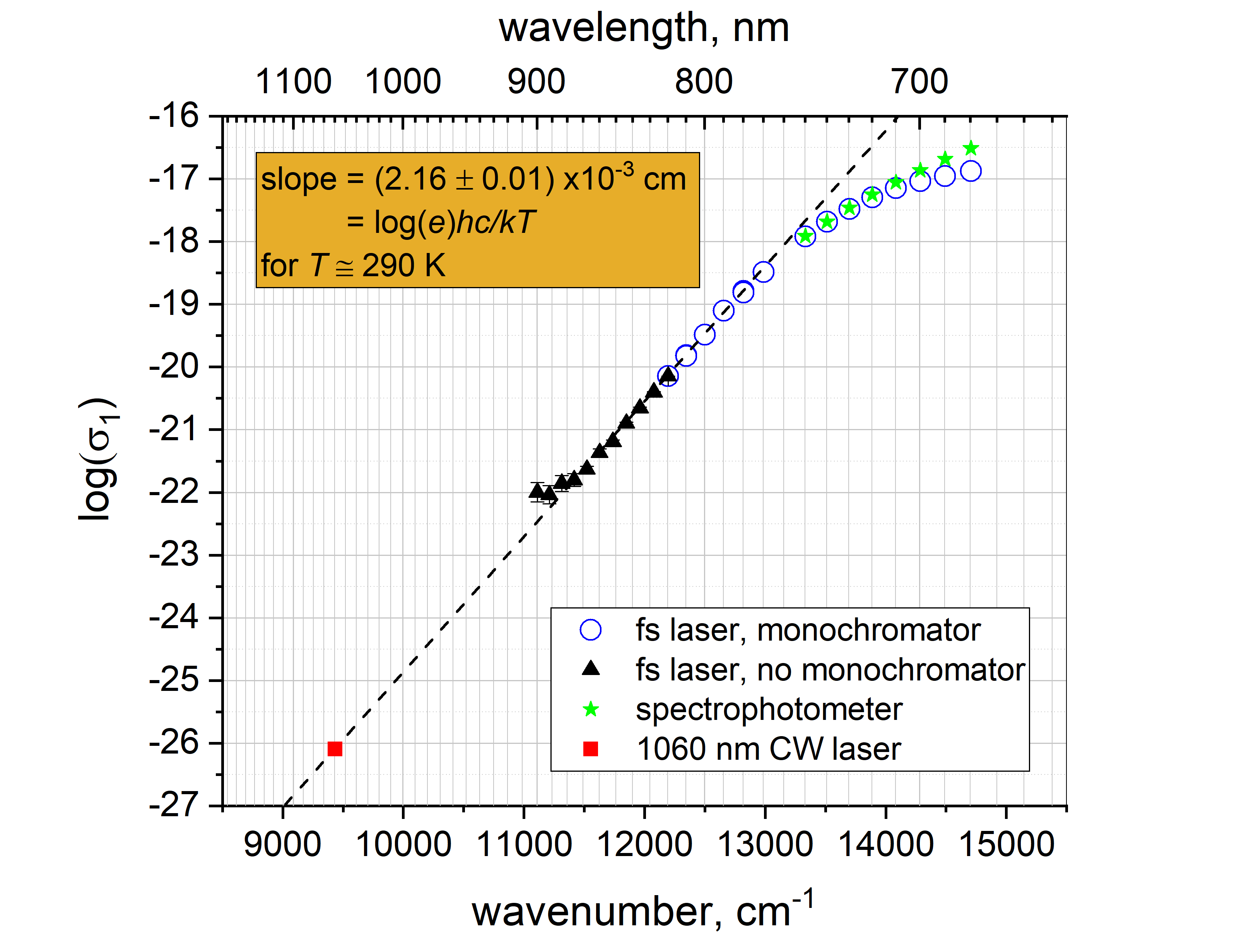}
     \caption{Experimental HBA cross section obtained using a tunable femtosecond laser (blue circles and black triangles), a spectrophotometer (green stars) and the 1060~nm CW laser (red square). The results lie along a linear fit (dashed line) with a slope in agreement with Eq.~\ref{SIsigmaHBA}.}
     \label{fig:1PAcrossSection}
\end{figure}

For the excitation wavelengths of 680 to 820~nm (blue circles) and 820 to 900~nm (black triangles), a setup (described previously~\cite{2020Drobizhev}) based on a tunable femtosecond laser coupled to a photon counting spectrofluorimeter was used. For the 680 to 820~nm range, the right channel of the spectrofluorimeter which contains a monochromator was used to collect fluorescence from the sample at 760~nm. The excitation power was varied from 0.4 to 40~mW and the resulting fluorescence signal was confirmed to depend linearly on excitation power. To derive a cross section from the signals, first, the ratio of the fluorescence signals at two independent wavelengths for the same excitation power is used to find a ratio of the HBA cross sections at the two wavelengths (see Eq.~1 (main text)). Next, the absolute value of the cross section is scaled in the 750 to 780~nm range to match the values reported in literature~\cite{2011Makarov}. The red tail of the absorption spectrum was also measured in a spectrophotometer (green stars) and scaled to match literature values.

For the 820 to 900~nm range, the left channel of the spectrofluorimeter was used to collect the integrated fluorescence signal through a 770~nm shortpass filter. Fluorescence was measured as a function of peak photon flux in the range $2\times10^{26}$ to $4\times10^{27}$~photons~cm$^{-2}$~s$^{-1}$. Similar to our results shown in Fig.~2~b (main text), as the photon flux is increased the signal transitions from depending linearly to quadratically on photon flux. At each wavelength, the signal as a function of flux was fit to Eq.~1 (main text). The coefficient from the linear term was used again to derive an HBA cross section. First, the relative strength was found using ratios of the coefficients at independent wavelengths as described above. Next, the absolute value is scaled at 820~nm to match the monochromator results. 

In the region of 770 to 876~nm, the cross sections obtained from these two methods are fit (black dashed line) to a linear function. The slope of the fit is found to equal $\mathrm{log}(e)hc/kT$ for $T\approx290$~K, which is consistent with the expected Boltzmann exponential dependence of the cross section (see Eq.~\ref{SIsigmaHBA}). The cross section also contains another frequency dependent term, the ratio of Franck-Condon factors (Fig.~\ref{fig:FrankCondon}~b), but here we assume these factors do not change significantly compared to the change imparted by the Boltzmann factor. 

For our 1060~nm CW laser measurement (red square), the signal from Fig.~2~b (main text) is fit to Eq.~1 (main text). The ratio of the coefficients in front of the linear and quadratic components is used to derive the HBA cross section, analogous to that shown in Eq.3 of Drobizhev \textit{et al.}~\cite{2003Drobizhev}, but for a CW beam with a Gaussian spatial profile. We used our derived value of $\sigma_{\text{C2PA}} = 220$~GM. Our derived HBA cross section lies along the fit to the HBA cross sections measured at lower wavelengths.

\bibliography{achemso-demo}

\providecommand{\latin}[1]{#1}
\makeatletter
\providecommand{\doi}
  {\begingroup\let\do\@makeother\dospecials
  \catcode`\{=1 \catcode`\}=2 \doi@aux}
\providecommand{\doi@aux}[1]{\endgroup\texttt{#1}}
\makeatother
\providecommand*\mcitethebibliography{\thebibliography}
\csname @ifundefined\endcsname{endmcitethebibliography}
  {\let\endmcitethebibliography\endthebibliography}{}
\begin{mcitethebibliography}{41}
\providecommand*\natexlab[1]{#1}
\providecommand*\mciteSetBstSublistMode[1]{}
\providecommand*\mciteSetBstMaxWidthForm[2]{}
\providecommand*\mciteBstWouldAddEndPuncttrue
  {\def\EndOfBibitem{\unskip.}}
\providecommand*\mciteBstWouldAddEndPunctfalse
  {\let\EndOfBibitem\relax}
\providecommand*\mciteSetBstMidEndSepPunct[3]{}
\providecommand*\mciteSetBstSublistLabelBeginEnd[3]{}
\providecommand*\EndOfBibitem{}
\mciteSetBstSublistMode{f}
\mciteSetBstMaxWidthForm{subitem}{(\alph{mcitesubitemcount})}
\mciteSetBstSublistLabelBeginEnd
  {\mcitemaxwidthsubitemform\space}
  {\relax}
  {\relax}

\bibitem[Gea-Banacloche(1989)]{1989Banacloche}
Gea-Banacloche,~J. {T}wo-Photon Absorption of Nonclassical Light. \emph{{P}hys.
  {R}ev. {L}ett.} \textbf{1989}, \emph{62}, 1603--1606\relax
\mciteBstWouldAddEndPuncttrue
\mciteSetBstMidEndSepPunct{\mcitedefaultmidpunct}
{\mcitedefaultendpunct}{\mcitedefaultseppunct}\relax
\EndOfBibitem
\bibitem[Javanianen and Gould(1990)Javanianen, and Gould]{1990Javanainen}
Javanianen,~J.; Gould,~P.~L. {L}inear Intensity Dependence of a Two-Photon
  Absorption Rate. \emph{{P}hys. {R}ev. {A}.} \textbf{1990}, \emph{41},
  5088--5091\relax
\mciteBstWouldAddEndPuncttrue
\mciteSetBstMidEndSepPunct{\mcitedefaultmidpunct}
{\mcitedefaultendpunct}{\mcitedefaultseppunct}\relax
\EndOfBibitem
\bibitem[Mollow(1968)]{1968Mollow}
Mollow,~B. {T}wo-Photon Absorption and Field Correlation Functions.
  \emph{{P}hys. {R}ev.} \textbf{1968}, \emph{175}, 1555--1563\relax
\mciteBstWouldAddEndPuncttrue
\mciteSetBstMidEndSepPunct{\mcitedefaultmidpunct}
{\mcitedefaultendpunct}{\mcitedefaultseppunct}\relax
\EndOfBibitem
\bibitem[Parzuchowski \latin{et~al.}(2021)Parzuchowski, Mikhaylov, Mazurek,
  Wilson, Lum, Jr., Gerrits, Stevens, and Jimenez]{2021Parzuchowski}
Parzuchowski,~K.~M.; Mikhaylov,~A.; Mazurek,~M.~D.; Wilson,~R.~N.; Lum,~D.~J.;
  Jr.,~C. H.~C.; Gerrits,~T.; Stevens,~M.~J.; Jimenez,~R. Setting Bounds on
  Entangled Two-Photon Absorption Cross Sections in Common Fluorophores.
  \emph{{P}hys. {R}ev. {A}ppl.} \textbf{2021}, \emph{15}, 044012\relax
\mciteBstWouldAddEndPuncttrue
\mciteSetBstMidEndSepPunct{\mcitedefaultmidpunct}
{\mcitedefaultendpunct}{\mcitedefaultseppunct}\relax
\EndOfBibitem
\bibitem[Raymer \latin{et~al.}(2020)Raymer, Landes, Allgaier, Merkouche, Smith,
  and Marcus]{2020Raymer}
Raymer,~M.~G.; Landes,~T.; Allgaier,~M.; Merkouche,~S.; Smith,~B.~J.;
  Marcus,~A.~H. Two-Photon Absorption of Time-Frequency-Entangled Photon Pairs
  by Molecules: The Roles of Photon-Number Correlations and Spectral
  Correlations. \emph{arXiv:2012.05375 [quant-ph]} \textbf{2020}, \relax
\mciteBstWouldAddEndPunctfalse
\mciteSetBstMidEndSepPunct{\mcitedefaultmidpunct}
{}{\mcitedefaultseppunct}\relax
\EndOfBibitem
\bibitem[Carnio \latin{et~al.}(2021)Carnio, Buchleitner, and
  Schlawin]{2021Carnio}
Carnio,~E.~G.; Buchleitner,~A.; Schlawin,~F. {H}ow To Optimize the Absorption
  of Two Entangled Photons. \emph{arXiv:2105.13876} \textbf{2021},
  \emph{[quant-ph]}\relax
\mciteBstWouldAddEndPuncttrue
\mciteSetBstMidEndSepPunct{\mcitedefaultmidpunct}
{\mcitedefaultendpunct}{\mcitedefaultseppunct}\relax
\EndOfBibitem
\bibitem[French and Goodson(2004)French, and Goodson]{2004French}
French,~R.~E.; Goodson,~T.~G. {A}pplications Of Correlated Photon Statistics
  With a Biphoton Source in an Organic Material. \emph{{P}roc. {SPIE}}
  \textbf{2004}, \emph{5161}\relax
\mciteBstWouldAddEndPuncttrue
\mciteSetBstMidEndSepPunct{\mcitedefaultmidpunct}
{\mcitedefaultendpunct}{\mcitedefaultseppunct}\relax
\EndOfBibitem
\bibitem[Lee and Goodson(2006)Lee, and Goodson]{2006Lee}
Lee,~D.-I.; Goodson,~T. {E}ntangled Photon Absorption in an Organic Porphyrin
  Dendrimer. \emph{{J}. {P}hys. {C}hem. B} \textbf{2006}, \emph{110},
  25582--25585\relax
\mciteBstWouldAddEndPuncttrue
\mciteSetBstMidEndSepPunct{\mcitedefaultmidpunct}
{\mcitedefaultendpunct}{\mcitedefaultseppunct}\relax
\EndOfBibitem
\bibitem[Harpham \latin{et~al.}(2009)Harpham, Suzer, Ma, Bauerle, and
  Goodson]{2009Harpham}
Harpham,~M.~R.; Suzer,~O.; Ma,~C.-Q.; Bauerle,~P.; Goodson,~T. {T}hiophene
  Dendrimers as Entangled Photon Sensor Materials. \emph{{J}. of {A}m. {C}hem.
  {S}oc.} \textbf{2009}, \emph{131}, 973--979\relax
\mciteBstWouldAddEndPuncttrue
\mciteSetBstMidEndSepPunct{\mcitedefaultmidpunct}
{\mcitedefaultendpunct}{\mcitedefaultseppunct}\relax
\EndOfBibitem
\bibitem[Guzman \latin{et~al.}(2010)Guzman, Harpham, Suzer, Haley, and
  Goodson]{2010Guzman}
Guzman,~A.~R.; Harpham,~M.~R.; Suzer,~O.; Haley,~M.~M.; Goodson,~T.~G.
  {S}patial Control of Entangled Two-Photon Absorption with Organic
  Chromophores. \emph{{J}. of {A}m. {C}hem. {S}oc.} \textbf{2010}, \emph{132},
  7840--7841\relax
\mciteBstWouldAddEndPuncttrue
\mciteSetBstMidEndSepPunct{\mcitedefaultmidpunct}
{\mcitedefaultendpunct}{\mcitedefaultseppunct}\relax
\EndOfBibitem
\bibitem[Upton \latin{et~al.}(2013)Upton, Harpham, Suzer, Richter, Mukamel, and
  Goodson]{2013Upton}
Upton,~L.; Harpham,~M.; Suzer,~O.; Richter,~M.; Mukamel,~S.; Goodson,~T.
  {O}ptically Excited Entangled States in Organic Molecules Illuminate the
  Dark. \emph{{J}. {P}hys. {C}hem. {L}ett.} \textbf{2013}, \emph{4},
  2046--2052\relax
\mciteBstWouldAddEndPuncttrue
\mciteSetBstMidEndSepPunct{\mcitedefaultmidpunct}
{\mcitedefaultendpunct}{\mcitedefaultseppunct}\relax
\EndOfBibitem
\bibitem[Varnavski \latin{et~al.}(2017)Varnavski, Pinsky, , and
  Goodson]{2017Varnavski}
Varnavski,~O.; Pinsky,~B.; ; Goodson,~T. {E}ntangled Photon Excited
  Fluorescence in Organic Materials: An Ultrafast Coincidence Detector.
  \emph{{J}. {P}hys. {C}hem. {L}ett.} \textbf{2017}, \emph{8}, 388--393\relax
\mciteBstWouldAddEndPuncttrue
\mciteSetBstMidEndSepPunct{\mcitedefaultmidpunct}
{\mcitedefaultendpunct}{\mcitedefaultseppunct}\relax
\EndOfBibitem
\bibitem[Villabona-Monsalve \latin{et~al.}(2018)Villabona-Monsalve, Varnavski,
  Palfey, and Goodson]{2018Monsalve}
Villabona-Monsalve,~J.~P.; Varnavski,~O.; Palfey,~B.~A.; Goodson,~T.
  {T}wo-Photon Excitation of Flavins and Flavoproteins with Classical and
  Quantum Light. \emph{{J}. {A}m. {C}hem. {S}oc.} \textbf{2018}, \emph{140},
  14562--14566\relax
\mciteBstWouldAddEndPuncttrue
\mciteSetBstMidEndSepPunct{\mcitedefaultmidpunct}
{\mcitedefaultendpunct}{\mcitedefaultseppunct}\relax
\EndOfBibitem
\bibitem[Kang \latin{et~al.}(2020)Kang, Avanaki, Mosquera, Burdick,
  Villabona-Monsalve, Goodson, and Schatz]{2020Schatz}
Kang,~G.; Avanaki,~K.~N.; Mosquera,~M.~A.; Burdick,~R.~K.;
  Villabona-Monsalve,~J.~P.; Goodson,~T.; Schatz,~G.~C. {E}fficient Modeling of
  Organic Chromophores for Entangled Two-Photon Absorption. \emph{{J}. of {A}m.
  {C}hem. {S}oc.} \textbf{2020}, \emph{142}, 10446--10458\relax
\mciteBstWouldAddEndPuncttrue
\mciteSetBstMidEndSepPunct{\mcitedefaultmidpunct}
{\mcitedefaultendpunct}{\mcitedefaultseppunct}\relax
\EndOfBibitem
\bibitem[Ashkenazy \latin{et~al.}(2019)Ashkenazy, Wang, Unternährer, Fixler,
  and Stefanov]{2019Ashkenazy}
Ashkenazy,~A.; Wang,~K.; Unternährer,~M.; Fixler,~D.; Stefanov,~A.
  {E}stimation Of the Rate of Entangled-Photon Pair Interaction With Metallic
  Nanoparticles Based on Classical-Light Second-Harmonic Generation
  Measurements. \emph{{J}. {P}hys. {B}: {A}t. {M}ol. {O}pt. {P}hys.}
  \textbf{2019}, \emph{52}, 145401--145411\relax
\mciteBstWouldAddEndPuncttrue
\mciteSetBstMidEndSepPunct{\mcitedefaultmidpunct}
{\mcitedefaultendpunct}{\mcitedefaultseppunct}\relax
\EndOfBibitem
\bibitem[Mikhaylov \latin{et~al.}(2021)Mikhaylov, Parzuchowski, Mazurek, Lum,
  Gerrits, Jr., Stevens, and Jimenez]{2020Mikhaylov}
Mikhaylov,~A.; Parzuchowski,~K.~M.; Mazurek,~M.~D.; Lum,~D.~J.; Gerrits,~T.;
  Jr.,~C. H.~C.; Stevens,~M.~J.; Jimenez,~R. {A} Comprehensive Experimental
  System for Measuring Molecular Two-Photon Absorption Using an Ultrafast
  Entangled Photon Pair Excitation Source. \emph{{P}roc. {SPIE}} \textbf{2021},
  \emph{11295}\relax
\mciteBstWouldAddEndPuncttrue
\mciteSetBstMidEndSepPunct{\mcitedefaultmidpunct}
{\mcitedefaultendpunct}{\mcitedefaultseppunct}\relax
\EndOfBibitem
\bibitem[Corona-Aquino \latin{et~al.}(2021)Corona-Aquino, Calderón-Losada,
  Li-Gómez, Cruz-Ramirez, Alvarez-Venicio, del Pilar Carreón-Castro,
  de~J.~León-Montiel, and U'Ren]{2021Corona}
Corona-Aquino,~S.; Calderón-Losada,~O.; Li-Gómez,~M.~Y.; Cruz-Ramirez,~H.;
  Alvarez-Venicio,~V.; del Pilar Carreón-Castro,~M.; de~J.~León-Montiel,~R.;
  U'Ren,~A.~B. {E}xperimental Study on the Effects of Photon-Pair Temporal
  Correlations in Entangled Two-Photon Absorption. \emph{arXiv:2101.10987}
  \textbf{2021}, \emph{[quant-ph]}\relax
\mciteBstWouldAddEndPuncttrue
\mciteSetBstMidEndSepPunct{\mcitedefaultmidpunct}
{\mcitedefaultendpunct}{\mcitedefaultseppunct}\relax
\EndOfBibitem
\bibitem[Raymer \latin{et~al.}(2021)Raymer, Landes, Allgaier, Merkouche, Smith,
  and Marcus]{2021Raymer}
Raymer,~M.~G.; Landes,~T.; Allgaier,~M.; Merkouche,~S.; Smith,~B.~J.;
  Marcus,~A.~H. {H}ow Large Is the Quantum Enhancement of Two-Photon Absorption
  by Time-Frequency Entanglement of Photon Pairs? \emph{{O}ptica}
  \textbf{2021}, \emph{8}, 757--758\relax
\mciteBstWouldAddEndPuncttrue
\mciteSetBstMidEndSepPunct{\mcitedefaultmidpunct}
{\mcitedefaultendpunct}{\mcitedefaultseppunct}\relax
\EndOfBibitem
\bibitem[Landes \latin{et~al.}(2021)Landes, Raymer, Allgaier, Merkouche, Smith,
  and Marcus]{2021Landes}
Landes,~T.; Raymer,~M.~G.; Allgaier,~M.; Merkouche,~S.; Smith,~B.~J.;
  Marcus,~A.~H. {Q}uantifying the Enhancement of Two-Photon Absorption Due to
  Spectral-Temporal Entanglement. \emph{{O}pt. {E}xp.} \textbf{2021},
  \emph{29}, 20022--20033\relax
\mciteBstWouldAddEndPuncttrue
\mciteSetBstMidEndSepPunct{\mcitedefaultmidpunct}
{\mcitedefaultendpunct}{\mcitedefaultseppunct}\relax
\EndOfBibitem
\bibitem[Tabakaev \latin{et~al.}(2021)Tabakaev, Montagnese, Haack, Bonacina,
  Wolf, Zbinden, and Thew]{2021Tabakaev}
Tabakaev,~D.; Montagnese,~M.; Haack,~G.; Bonacina,~L.; Wolf,~J.-P.;
  Zbinden,~H.; Thew,~R.~T. Energy-Time-Entangled Two-Photon Molecular
  Absorption. \emph{{P}hys. {R}ev. A} \textbf{2021}, \emph{103}, 033701\relax
\mciteBstWouldAddEndPuncttrue
\mciteSetBstMidEndSepPunct{\mcitedefaultmidpunct}
{\mcitedefaultendpunct}{\mcitedefaultseppunct}\relax
\EndOfBibitem
\bibitem[Landes \latin{et~al.}(2021)Landes, Allgaier, Merkouche, Smith, Marcus,
  and Raymer]{2020landes}
Landes,~T.; Allgaier,~M.; Merkouche,~S.; Smith,~B.~J.; Marcus,~A.~H.;
  Raymer,~M.~G. Experimental feasibility of molecular two-photon absorption
  with isolated time-frequency-entangled photon pairs. \emph{Phys. Rev.
  Research} \textbf{2021}, \emph{3}, 033154\relax
\mciteBstWouldAddEndPuncttrue
\mciteSetBstMidEndSepPunct{\mcitedefaultmidpunct}
{\mcitedefaultendpunct}{\mcitedefaultseppunct}\relax
\EndOfBibitem
\bibitem[Lavoie \latin{et~al.}(2020)Lavoie, Landes, Tamimi, Smith, Marcus, and
  Raymer]{2020Lavoie}
Lavoie,~J.; Landes,~T.; Tamimi,~A.; Smith,~B.~J.; Marcus,~A.~H.; Raymer,~M.~G.
  {P}hase-Modulated Interferometry, Spectroscopy, and Refractometry using
  Entangled Photon Pairs. \emph{{A}dv. {Q}uan. {T}echn.} \textbf{2020},
  \emph{3}, 1900114\relax
\mciteBstWouldAddEndPuncttrue
\mciteSetBstMidEndSepPunct{\mcitedefaultmidpunct}
{\mcitedefaultendpunct}{\mcitedefaultseppunct}\relax
\EndOfBibitem
\bibitem[Apatin and Makarov(1983)Apatin, and Makarov]{1983Apatin}
Apatin,~V.; Makarov,~G. {M}ultiphoton Absorption of Infrared Laser Radiation by
  $\text{SF}_6$ Molecules Cooled in a Supersonic Jet. \emph{{Z}h. {E}ksp.
  {T}eor. {F}iz.} \textbf{1983}, \emph{84}, 15--29\relax
\mciteBstWouldAddEndPuncttrue
\mciteSetBstMidEndSepPunct{\mcitedefaultmidpunct}
{\mcitedefaultendpunct}{\mcitedefaultseppunct}\relax
\EndOfBibitem
\bibitem[Okamura \latin{et~al.}(1996)Okamura, Tosa, Ishii, and
  Takeuchi]{1995Okamura}
Okamura,~H.; Tosa,~V.; Ishii,~T.; Takeuchi,~K. {C}ollisional Effects in the
  {IR} Multiphoton Absorption and Dissociation Of $\text{Si}_2\text{F}_6$.
  \emph{{J}. of {P}hotochem. and {P}hotobiol. {A}:{C}hem.} \textbf{1996},
  \emph{95}, 203--207\relax
\mciteBstWouldAddEndPuncttrue
\mciteSetBstMidEndSepPunct{\mcitedefaultmidpunct}
{\mcitedefaultendpunct}{\mcitedefaultseppunct}\relax
\EndOfBibitem
\bibitem[Drobizhev \latin{et~al.}(2003)Drobizhev, Karotki, Kruk, Krivokapic,
  Anderson, and Rebane]{2003Drobizhev}
Drobizhev,~M.; Karotki,~A.; Kruk,~M.; Krivokapic,~A.; Anderson,~H.~L.;
  Rebane,~A. {P}hoton Energy Upconversion in Porphyrins: One-Photon Hot-Band
  Absorption Versus Two-Photon Absorption. \emph{{C}hem.~{P}hys. {L}ett.}
  \textbf{2003}, \emph{370}, 690--699\relax
\mciteBstWouldAddEndPuncttrue
\mciteSetBstMidEndSepPunct{\mcitedefaultmidpunct}
{\mcitedefaultendpunct}{\mcitedefaultseppunct}\relax
\EndOfBibitem
\bibitem[Makarov \latin{et~al.}(2007)Makarov, Drobizhev, Rebane, Peone, Wolleb,
  Spahni, Makarova, and Lukyanets]{2007Makarov}
Makarov,~N.; Drobizhev,~M.; Rebane,~A.; Peone,~D.; Wolleb,~H.; Spahni,~H.;
  Makarova,~E.; Lukyanets,~E. {R}esonance Enhancement of Two-Photon Absorption
  of Phthalocyanines for {3D} Optical Storage in the Presence of Hot-Band
  Absorption. \emph{{P}roc. {SPIE}} \textbf{2007}, \emph{6470}\relax
\mciteBstWouldAddEndPuncttrue
\mciteSetBstMidEndSepPunct{\mcitedefaultmidpunct}
{\mcitedefaultendpunct}{\mcitedefaultseppunct}\relax
\EndOfBibitem
\bibitem[Rebane \latin{et~al.}(2007)Rebane, Makarov, Drobizhev, Spangler, Gong,
  and Meng]{2007Rebane}
Rebane,~A.; Makarov,~N.; Drobizhev,~M.; Spangler,~C.~W.; Gong,~A.; Meng,~F.
  {B}road Bandwidth Near-{IR} Two-Photon Absorption in Conjugated
  Porphyrin-Core Dendrimers. \emph{{P}roc. {SPIE}} \textbf{2007},
  \emph{6653}\relax
\mciteBstWouldAddEndPuncttrue
\mciteSetBstMidEndSepPunct{\mcitedefaultmidpunct}
{\mcitedefaultendpunct}{\mcitedefaultseppunct}\relax
\EndOfBibitem
\bibitem[Starkey \latin{et~al.}(2008)Starkey, Rebane, Drobizhev, Meng, Gong,
  Elliott, McInnerney, and Spangler]{2008Starkey}
Starkey,~J.~R.; Rebane,~A.~K.; Drobizhev,~M.~A.; Meng,~F.; Gong,~A.;
  Elliott,~A.; McInnerney,~K.; Spangler,~C.~W. {N}ew Two-Photon Activated
  Photodynamic Therapy Sensitizers Induce Xenograft Tumor Regressions After
  Near-IR Laser Treatment Through the Body of the Host Mouse. \emph{{C}lin.
  {C}ancer {R}es.} \textbf{2008}, \emph{14}, 6564--6573\relax
\mciteBstWouldAddEndPuncttrue
\mciteSetBstMidEndSepPunct{\mcitedefaultmidpunct}
{\mcitedefaultendpunct}{\mcitedefaultseppunct}\relax
\EndOfBibitem
\bibitem[Dayan \latin{et~al.}(2005)Dayan, Pe'er, Friesem, and
  Silbernerg]{2005Dayan}
Dayan,~B.; Pe'er,~A.; Friesem,~A.~A.; Silbernerg,~Y. {N}onlinear Interactions
  With an Ultrahigh Flux of Broadband Entangled Photons. \emph{{P}hys. {R}ev.
  {L}ett.} \textbf{2005}, \emph{94}, 043602\relax
\mciteBstWouldAddEndPuncttrue
\mciteSetBstMidEndSepPunct{\mcitedefaultmidpunct}
{\mcitedefaultendpunct}{\mcitedefaultseppunct}\relax
\EndOfBibitem
\bibitem[de~Reguardati \latin{et~al.}(2016)de~Reguardati, Pahapill, Mikhailov,
  Stepanenko, and Rebane]{2016deReguardatti}
de~Reguardati,~S.; Pahapill,~J.; Mikhailov,~A.; Stepanenko,~Y.; Rebane,~A.
  High-Accuracy Reference Standards for Two-Photon Absorption in the 680–1050
  nm Wavelength Range. \emph{{O}pt. {E}xp.} \textbf{2016}, \emph{24},
  9053--9066\relax
\mciteBstWouldAddEndPuncttrue
\mciteSetBstMidEndSepPunct{\mcitedefaultmidpunct}
{\mcitedefaultendpunct}{\mcitedefaultseppunct}\relax
\EndOfBibitem
\bibitem[Makarov \latin{et~al.}(2011)Makarov, Campo, Hales, and
  Perry]{2011Makarov}
Makarov,~N.~S.; Campo,~J.; Hales,~J.~M.; Perry,~J.~W. {R}apid, Broadband
  Two-Photon-Excited Fluorescence Spectroscopy and its Application to
  Red-Emitting Secondary Reference Compounds. \emph{{O}pt.~{M}at. {E}xp.}
  \textbf{2011}, \emph{1}, 551--563\relax
\mciteBstWouldAddEndPuncttrue
\mciteSetBstMidEndSepPunct{\mcitedefaultmidpunct}
{\mcitedefaultendpunct}{\mcitedefaultseppunct}\relax
\EndOfBibitem
\bibitem[Drobizhev \latin{et~al.}(2020)Drobizhev, Molina, and
  Hughes]{2020Drobizhev}
Drobizhev,~M.; Molina,~R.~S.; Hughes,~T.~E. {C}haracterizing the Two-Photon
  Absorption Properties of Fluorescent Molecules in the 680–1300~nm Spectral
  Range. \emph{{B}io {P}rotoc.} \textbf{2020}, \emph{10}, 1--47\relax
\mciteBstWouldAddEndPuncttrue
\mciteSetBstMidEndSepPunct{\mcitedefaultmidpunct}
{\mcitedefaultendpunct}{\mcitedefaultseppunct}\relax
\EndOfBibitem
\bibitem[Fei \latin{et~al.}(1997)Fei, Jost, Popescu, Saleh, and Teich]{1997Fei}
Fei,~H.-B.; Jost,~B.~M.; Popescu,~S.; Saleh,~B. E.~A.; Teich,~M.
  {E}ntangled-Induced Two-Photon Transparency. \emph{{P}hys. {R}ev. {L}ett.}
  \textbf{1997}, \emph{78}, 1679--1682\relax
\mciteBstWouldAddEndPuncttrue
\mciteSetBstMidEndSepPunct{\mcitedefaultmidpunct}
{\mcitedefaultendpunct}{\mcitedefaultseppunct}\relax
\EndOfBibitem
\bibitem[Penzkofer and Leupacher(1987)Penzkofer, and Leupacher]{1987Penzkofer}
Penzkofer,~A.; Leupacher,~W. {F}luorescence Behaviour of Highly Concentrated
  Rhodamine 6G Solutions. \emph{{J}. of {L}um.} \textbf{1987}, \emph{37},
  61--72\relax
\mciteBstWouldAddEndPuncttrue
\mciteSetBstMidEndSepPunct{\mcitedefaultmidpunct}
{\mcitedefaultendpunct}{\mcitedefaultseppunct}\relax
\EndOfBibitem
\bibitem[Doan \latin{et~al.}(2017)Doan, Castillo, Bejjani, Nurekeyev, Dzyuba,
  Gryczynski, Gryczynski, and Raut]{Doan2017}
Doan,~H.; Castillo,~M.; Bejjani,~M.; Nurekeyev,~Z.; Dzyuba,~S.~V.;
  Gryczynski,~I.; Gryczynski,~Z.; Raut,~S. Solvatochromic Dye LDS 798 As
  Microviscosity and pH Probe. \emph{Phys. Chem. Chem. Phys.} \textbf{2017},
  \emph{19}, 29934--29939\relax
\mciteBstWouldAddEndPuncttrue
\mciteSetBstMidEndSepPunct{\mcitedefaultmidpunct}
{\mcitedefaultendpunct}{\mcitedefaultseppunct}\relax
\EndOfBibitem
\bibitem[Luchowski \latin{et~al.}(2009)Luchowski, Gryczynski, Sarkar, Borejdo,
  Szabelski, Kapusta, and Gryczynski]{2009Luchowski}
Luchowski,~R.; Gryczynski,~Z.; Sarkar,~P.; Borejdo,~J.; Szabelski,~M.;
  Kapusta,~P.; Gryczynski,~I. {I}nstrument Response Standard in Time-Resolved
  Fluorescence. \emph{{R}ev. {S}ci. {I}nstrum.} \textbf{2009}, \emph{80},
  033109\relax
\mciteBstWouldAddEndPuncttrue
\mciteSetBstMidEndSepPunct{\mcitedefaultmidpunct}
{\mcitedefaultendpunct}{\mcitedefaultseppunct}\relax
\EndOfBibitem
\bibitem[Kachynski \latin{et~al.}(2005)Kachynski, Kuzmin, Pudavar, and
  Prasad]{2005Kachynski}
Kachynski,~A.; Kuzmin,~A.; Pudavar,~H.; Prasad,~P. {T}hree-Dimensional Confocal
  Thermal Imaging Using Anti-{S}tokes Luminescence. \emph{{A}ppl.~{P}hys.
  {L}ett.} \textbf{2005}, \emph{87}, 023901\relax
\mciteBstWouldAddEndPuncttrue
\mciteSetBstMidEndSepPunct{\mcitedefaultmidpunct}
{\mcitedefaultendpunct}{\mcitedefaultseppunct}\relax
\EndOfBibitem
\bibitem[Lakowicz(2006)]{2006Lakowicz}
Lakowicz,~J.~R. \emph{{P}rinciples of {F}luorescence {S}pectroscopy}, 3rd ed.;
  Springer: New York, NY, USA, 2006\relax
\mciteBstWouldAddEndPuncttrue
\mciteSetBstMidEndSepPunct{\mcitedefaultmidpunct}
{\mcitedefaultendpunct}{\mcitedefaultseppunct}\relax
\EndOfBibitem
\bibitem[Lerch \latin{et~al.}(2013)Lerch, Bessire, Bernhard, Feurer, and
  Stefanov]{2013Lerch}
Lerch,~S.; Bessire,~B.; Bernhard,~C.; Feurer,~T.; Stefanov,~A. {T}uning Curve
  of Type-0 Spontaneous Parametric Down-Conversion. \emph{{J}. of {O}pt. {S}oc.
  of {A}m.} \textbf{2013}, \emph{30}, 953--958\relax
\mciteBstWouldAddEndPuncttrue
\mciteSetBstMidEndSepPunct{\mcitedefaultmidpunct}
{\mcitedefaultendpunct}{\mcitedefaultseppunct}\relax
\EndOfBibitem
\bibitem[Szoke \latin{et~al.}(2021)Szoke, He, Hickam, and Cushing]{2021Szoke}
Szoke,~S.; He,~M.; Hickam,~B.; Cushing,~S. {D}esigning High-Power, Octave
  Spanning Entangled Photon Sources for Quantum Spectroscopy. \emph{{J}.
  {C}hem. {P}hys.} \textbf{2021}, \emph{154}, 244201\relax
\mciteBstWouldAddEndPuncttrue
\mciteSetBstMidEndSepPunct{\mcitedefaultmidpunct}
{\mcitedefaultendpunct}{\mcitedefaultseppunct}\relax
\EndOfBibitem
\end{mcitethebibliography}

\end{suppinfo}
\end{document}